\documentclass{jfm}
\usepackage{graphicx} 
\usepackage{subcaption}
\usepackage{titlesec}
\usepackage{xcolor}
\usepackage{verbatim}
\usepackage{amssymb}
\usepackage{amsmath}
\usepackage{ifthen}
\usepackage{enumitem}
\usepackage{cancel}
\usepackage{ulem}
\usepackage{cite}
\usepackage{times}
\usepackage{setspace}
\usepackage{wrapfig}
\usepackage{bm}
\usepackage{hyperref}
\newboolean{showcomments}
\setboolean{showcomments}{true}

\newcommand{\be}{\begin{equation}}
\newcommand{\ee}{\end{equation}}
\newcommand{\filelinknotebook}{ 
\href{https://www.cambridge.org/core/journals/journal-of-fluid-mechanics/jfm-notebooks}{\color{blue}https://cocalc.com/Cambridge/JFM-preprint-notebook}}

\definecolor{bl}{rgb}{0.,0.1,0.75}
\newcommand\dc[1]{{\color{blue} #1}}
\newcommand\nhu[1]{{\color{magenta} #1}}

\shorttitle{Wind-shade roughness model}
\shortauthor{C. Meneveau, N. Hutchins \& D. Chung}

\title{The wind-shade roughness model for turbulent wall-bounded flows}

\author{Charles Meneveau\aff{1}\corresp{\email{meneveau@jhu.edu}},
   Nicholas Hutchins\aff{2} \& Daniel Chung\aff{2}
   }

\affiliation{\aff{1}Department of Mechanical Engineering, Johns Hopkins University
\aff{2}Department of Mechanical Engineering, University of Melbourne
}

\begin{document}
 
\maketitle

\begin{abstract}
To aid in prediction of turbulent boundary layer flows over rough surfaces, a new model is proposed to estimate hydrodynamic roughness based solely on geometric surface information. The model is based on a fluid-mechanics motivated geometric parameter called the wind-shade factor.  Sheltering is included using a rapid algorithm adapted from the landscape shadow literature, while local pressure drag is estimated using a piecewise potential flow approximation. Similarly to evaluating traditional surface parameters such as skewness or average slope magnitude, the wind-shade factor is purely geometric and can be evaluated efficiently from knowing the surface elevation map and the mean flow direction. The wind-shade roughness model is applied to over 100 different surfaces available in a public roughness database and some others, and the predicted sandgrain-roughness heights are compared to measured values.  Effects of various model ingredients are analyzed, and transitionally rough surfaces are treated by adding  a term representing the viscous stress component. 

\end{abstract}

\section{Introduction}

Surface roughness profoundly impacts turbulent boundary layers.
Relative to a smooth surface, roughness increases not only drag but also heat and mass transfer, with consequences on efficiency, emissions and the climate.
In international shipping, for example, biofouled-roughened ship hulls increase fuel consumption by tens of billions of dollars annually,
along with a proportional increase in greenhouse gas emissions.
Yet our ability to manage the consequences is paced by our skill in predicting the drag of rough surfaces, a longstanding problem in fluid mechanics \citep{Raupach1991,Jimenez2004,Flack2010,chung2021predicting}.

Roughness effects cannot be easily generalised.
Barnacles on ship hulls do not engender drag in the same way as trees in the atmospheric surface layer.
However, the opposite extreme view, that roughness effects cannot be generalised at all, is also unfounded.
Intuitively, the larger the roughness size, the greater the drag.
And, for a given size,    it has long been appreciated that the roughness plan density, the ratio of roughness plan area to wall area, plays a key role in determining drag \citep{Schlichting1937}.
This knowledge is subsequently codified in models and frequently refined as new data became available to address model weaknesses \citep{flack2022important}.

Roughness models are formulas for drag that are fitted to topographical parameters.
Topographical parameters only depend on the roughness geometry.
In this way, models link drag with roughness features that are likely to matter.
For example, the roughness frontal density is important because pressure drag is proportional to the frontal area \citep{simpson1973generalized}.
Skewness captures the observation that peaks are more draggy than valleys \citep{Flack2010,jelly2018reynolds} and effective slope captures the observation that steeper slopes are more draggy than shallow slopes \citep{napoli2008effect}.
As it became clear which sets of these parameters are independent \citep{placidi2015,thakkar2017surface}, and with ready access to rapid prototyping both in laboratory experiments (computer numerical control machining) and in numerical simulations (immersed boundaries), research shifted to systematic sweeps in parameter space \citep{Schultz1999,chan2015systematic,forooghi2017toward,barros2018measurements,kuwata2019direct,flack2020skin,ma2020scaling,jouybari2021data,yang2022direct,jelly2022impact}.
Machine learning tools have now been brought to bear on the rapidly growing dataset.
To account for the many types of surfaces, more extensive statistical features \citep{jouybari2021data}, and even the surface-elevation probability density functions and power-spectral densities are taken as input \citep{yang2023prediction,yousefi2024machine}.
As the roughness parameter space is infinite, with unfamiliar surfaces or unexpected behaviour reported from time to time \citep{nugroho2021non,barros2018measurements,hutchins2023defining,womack2022turbulent}, physically interpretable predictions are essential for reliability \citep{brunton2020machine}, but how to do so is an active area of research.

One approach to physical interpretability is to use topographical parameters motivated by or derived from flow physics.
In addition to interpretability, topographical parameters are relatively easy to port, e.g.\ in a dynamic procedure of a large-eddy simulation \citep{anderson2011dynamic} or in a neural network.
A benefit of interpretability is the underlying physical hypothesis can be tested and advanced.
Ideally only a few parameters and a few fitting constants are used.
One enduring parameter is the frontal solidity \citep{Schlichting1937,Jimenez2004} or closely related parameters such as the effective slope.
When roughness features are sparsely spaced, the higher the frontal solidity, the higher the drag.
However, when roughness features are densely packed to shelter one another from the oncoming flow, the higher the frontal solidity, the lower the drag.
This sheltering effect is described comprehensively by \citet{Grimmond1999} and encapsulated in the formula of \citet{Macdonald1998}, whose predictive skill was found to be impressive in untested parameter spaces \citep{chung2021predicting}.
Sheltering was extended in a series of studies \citep{yang2016exponential,sadique2017aerodynamic,yang2016large,yang2017modelling} to directly predict drag from surface elevation maps of tall prisms, fractal and directional surfaces, bypassing the use of topographical parameters.
In addition to regular surfaces, sheltering also applies to irregular surfaces \citep{yang2022direct} and it turns out that sheltering also plays a decisive role on heat transfer \citep{rowin2024modelling}.
Although frontal solidity was used and discussed hand-in-hand with sheltering, by itself the frontal solidity does not encapsulate our current understanding of roughness flow physics.
For example, the frontal solidity does not discriminate between the differences in pressure drag due to gentle versus steep roughness slopes.
A related point is that the frontal solidity needs to be used with the plan solidity or skewness to differentiate sparse surfaces with occasional spikes from wavy surfaces.
Although the link between pressure drag and frontal area is proportional, the addition of plan solidity or skewness is somewhat adhoc. In addition, frontal solidity, even with the inclusion of skewness or plan solidity, cannot account for clustering, which can alter levels of sheltering of roughness elements \citep{sarakinos2022investigation}.

The present effort is focused on a physics-based analysis to determine and propose a geometric parameter, called the wind shade roughness factor, which by itself has predictive capabilities. In the future, it can also be combined with other parameters and further extend possible machine learning approaches, or, more simply, be used as single parameter. The proposed predictive model based on the wind-shade roughness factor is easily implemented by other researchers, and to further ease of implementation, the code and some example applications are provided as a computational notebook together with this paper. The intended use for the present   roughness model is routine calculations based on available elevation maps, reduced to a simple geometric parameter like the skewness or effective slope, without involving complicated solutions to non-local partial differential equations (e.g.\ solving Laplace's equation) such as the force-partitioning-inspired method \citep{aghaei2022contributions} or having to perform fluid dynamics simulations.
Moreover, the number of fitting constants required to reproduce data should be much smaller than the number of data points available for validation.

In the present paper, the wind-shade factor can be seen as a geometric parameter that combines both effects of plan and frontal solidities, or skewness and effective slope.
Its physical derivation is presented in \S\ref{sec:windshademodel} while the effects of viscosity for transitional roughness are added in \S\ref{sec:transitional}.
Predictions and comparisons with data are given in \S\ref{sec:results}, and the impact of various modelling choices is analysed in \S\ref{sec:analysis}. Concluding remarks are provided in \S \ref{sec:conclusions}.

\section{The wind-shade roughness model}\label{sec:windshademodel}

We consider a rough surface whose elevation is given by a single-valued  function $h(x,y)$. The first goal is to estimate the form (pressure) drag force arising when this surface is placed in a turbulent flow, under fully rough conditions.  For such conditions, the  representation should depend only upon the surface geometry,  the direction of the flow with respect to that geometry (e.g. an angle $\phi$) and the turbulence spreading angle $\theta$. The overall tangential force $f_i$ ($i=1,2$) (per unit mass) on the rough wall is expressed as an area integral of the pressure-caused kinematic wall stress along the entire surface:
\be
f_i = - \iint_A \tau^p_{i3}(x,y) \,\, dx dy = - \iint_A \frac{1}{\rho} p(x,y) \, \frac{\partial h}{\partial x_i}  \, F^{\rm sh}(x,y;\phi,\theta) \, dx dy.
\label{eq:figeneral}
\ee
The pressure field $p(x,y)$ is relative to some reference pressure $p_\infty=0$. The factor $F^{\rm sh}(x,y;\phi,\theta)$ is the sheltering factor described in the next section.  

\subsection{Sheltering function}
The sheltering function $F^{\rm sh}(x,y;\theta)$ is defined as 
\be F^{\rm sh}(x,y;\theta)=H(\theta - \beta(x,y)),
\label{eq:defshadefunc}
\ee
where $\beta(x,y)$ is the angle made by any given point with its upstream horizon, while $\theta$ is the turbulence spreading angle. $H(x)$ is the Heaviside function.  The sheltering function requires knowledge of the angle $\beta(x,y)$, the angle between the horizontal direction and a point's ``backward horizon'' point \citep{dozier1981faster}. To define the backward horizon function, it is convenient to assume that the $x$ direction is aligned with the incoming velocity $u_i$, i.e. $u_2=0$  (any dependence on $\phi$ omitted henceforth from the notation). The function $x_b(x)$ is called the ``backward horizon function'' of point $x$ \citep{dozier1981faster} and is the position of the visible horizon looking into the negative direction (or upstream towards the incoming flow) from point $x$.  For the maximum of $h(x,y)$, one sets $x_b(x)=x$.  
Then the horizon angle at point $(x,y)$ is given by  
\be \tan \beta = \frac{h(x_b(x),y)-h(x,y)}{x_b(x)-x}.\ee 
 
\begin{figure}
  \centerline{\includegraphics[width=8cm]{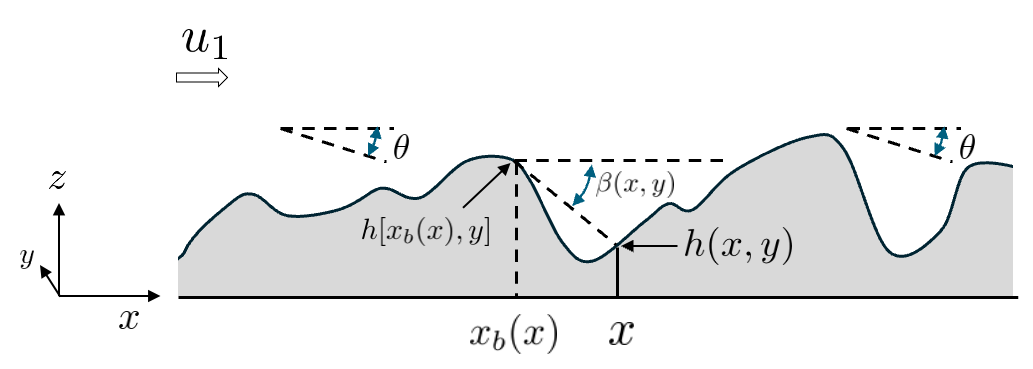}}
  \vspace{-0.1in}
  \caption{Sketch of surface, turbulence spreading angle $\theta$ , and  backward horizon angle $\beta(x,y)$ for any given point $(x,y)$. Since in the example shown the point $x$ has a backward horizon angle $\beta$ that is larger than the turbulence spreading angle $\theta$, point $x$ is considered sheltered (wind shaded). }
   \vspace{-0.1in}
\label{fig:sketchangles}
\end{figure}

Figure \ref{fig:sketchangles} shows the various angles for any given position $x$ as well as its horizon function $x_b(x)$. The angle distribution $\beta(x,y)$ depends only upon the surface geometry via the backward horizon function and need only be determined once for a given surface (i.e. it does not depend upon the turbulence spreading angle $\theta$). The landscape shading literature (e.g. \cite{dozier1981faster}) includes fast,  $O(N)$, algorithms to determine the horizon function $x_b(x)$. 
With these methods, evaluation of the sheltering factor $F^{\rm sh}(x,y;\theta)$ can be done very efficiently for any given $\theta$ since $\beta(x,y)$ can be pre-computed for a given surface.

\begin{figure}
    \centering
\includegraphics[width=1.0\linewidth]{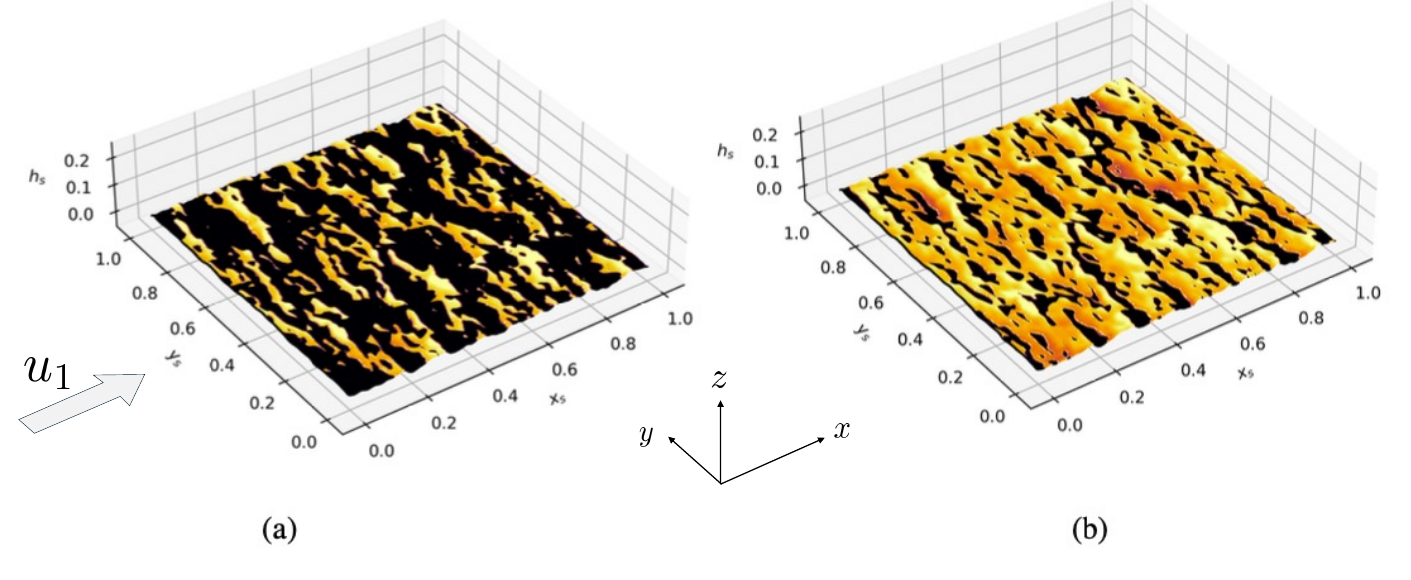}
\vspace{-0.1in}
\caption{Surface and its shaded regions (in black) for two angles $\theta=5$ degrees (a) and $\theta=15$ degrees (b). Computation of shaded region is based on the fast algorithm of \cite{dozier1981faster}, separately for each $x$-line in the direction of the incoming flow velocity $u_1$. In this and all subsequent elevation map visualizations, in order to represent the appropriate relative dimensions and surface slopes, the axes are normalized by a single scale $L_y$, the width  of the domain, i.e., $x_s=x/L_y$,  $y_s=y/L_y$ and  $h_s=h/L_y$. }
\label{fig:sampleshaded}
\end{figure}

As illustration, in Fig. \ref{fig:sampleshaded} we show sample surface (case of turbine-type roughness with  transverse ($y$-aligned)  ridges from   \cite{jelly2022impact}) for two angles: $\theta=5$ degrees (left) and $\theta=15$ degrees (right). Black regions are shaded regions. The drag (and hence the effective hydrodynamic roughness height for case (b) is expected to be larger due to more frontal area being exposed to the flow. 

\subsection{Modeled pressure distribution}

To determine the pressure distribution $p(x,y)$, it is useful select a reference velocity. We chose a notional velocity denoted by $U_k$ representing the velocity at a height $z_k$ above the mean surface elevation. The height $z_k$ is assumed to be sufficiently large so that the mean velocity there can be assumed to be independent of $(x,y)$. We thus aim for a height just above the roughness sublayer where assuming that the velocity is constant over $(x,y)$ is appropriate. 

\begin{figure}
  \centerline{\includegraphics[width=14cm]{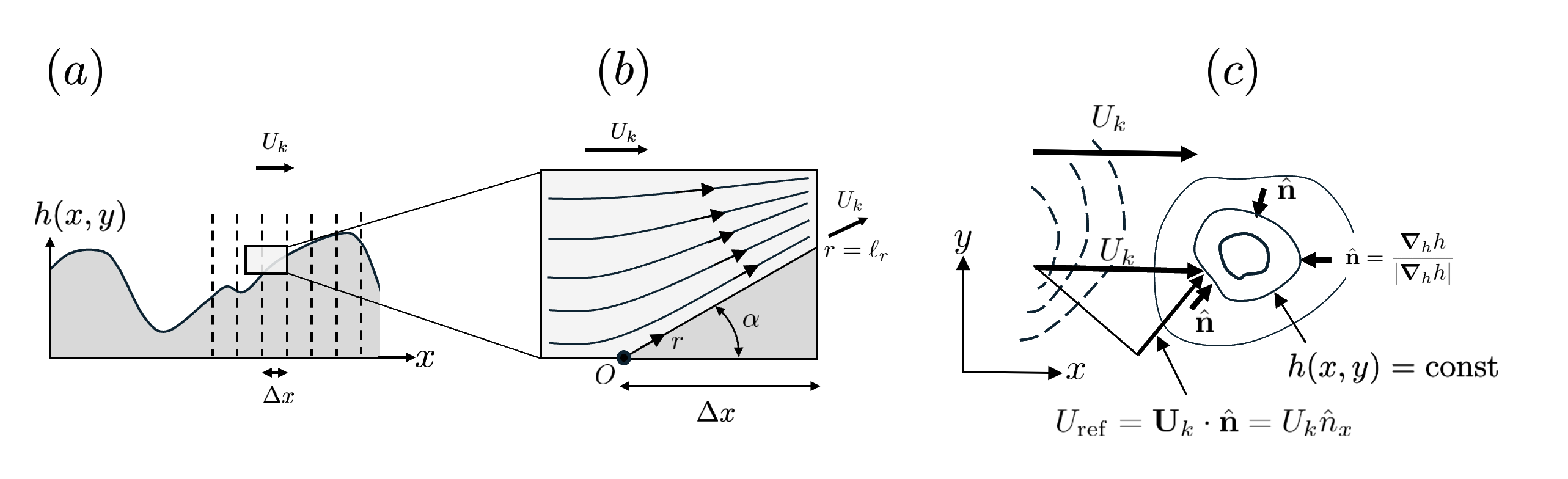}}
  \vspace{-0.1in}
  \caption{Sketch of surface discretized at horizontal resolution $\Delta x$ and (b) potential flow over a ramp at angle $\alpha$ \citep{ayala2024moving} assumed to be locally valid over the surface at horizontal discretization length $\Delta x$. In (a) and (b) the surface slope is assumed to be only in the $x$ direction, i.e. $\hat{n}_x=1$. The lowest point of the ramp has a stagnation point while at the end point the velocity magnitude is assumed to be $U_{\rm ref}$ and pressure is $p_\infty=0$. (c) shows a sketch of isosurface contours (surface seen from above), the local normal vector $\hat{\bf n} = \boldsymbol{\nabla}_h h/|\boldsymbol{\nabla}_h h|$,  the incoming velocity $U_k$ in the $x$ direction, and the incoming velocity normal to the surface becomes $U_{\rm ref} = U_k \hat{n}_x$.}
   \vspace{-0.1in}
\label{fig:sketchrampflow}
\end{figure}

We now turn to the pressure (form) drag model. We use the piece-wise potential ramp flow approach \citep{ayala2024moving} (see Fig. \ref{fig:sketchrampflow}) to model $p(x,y)$. We simplify any sloping portion of a roughness element or portions of a surface as a planar ramp inclined at an angle $\alpha$. The ramp angle is obtained from $\alpha = \arctan |\boldsymbol{\nabla}_h h|$ (i.e. based on the absolute value of the surface slope, for reasons to be explained below), and 
$\boldsymbol{\nabla}_h$ is the gradient in the horizontal plane, $\boldsymbol{\nabla}_h=\partial_x {\bf i}+\partial_y {\bf j}$. We assume at the bottom of each ramp flow, there is a stagnation point (solid circle in Fig. 
\ref{fig:sketchrampflow}b)
and at the top of the ramp the velocity is $U_k$ and the pressure there is equal to the reference pressure $p_\infty=0$, i.e. we assume plug flow between height $z_k$ and the ramp top. Evidently this is a strong assumption but is necessary if we wish to apply a potential flow description that is purely ``local'', i.e. is  agnostic of flow conditions at other locations and far above the surface.  

Following the development in \cite{ayala2024moving}, we use the known solution for potential flow over a ramp, for which the streamfunction in polar coordinates is given by $\psi(r,\theta) = A r^n \sin[n(\theta-\alpha)]$ with $n=\pi/(\pi-\alpha)$ and $A$ a constant. The radial coordinate runs from $r=0$ at the stagnation point to $r={\ell}_r$ at the top of the ramp,
where the velocity is a known reference velocity $U_{\rm ref}$ and the pressure is $p_\infty=0$. The radial (tangential to the surface) velocity component along the ramp surface is $V_r=-(1/r)\partial \psi/\partial \theta$ evaluated at $\theta=\alpha$. 
It results in $V_r(r) = n A \, (r/\ell_r)^{n-1}$,
which can be seen to fix $n A=U_{\rm ref}$. Using Bernoulli equation to evaluate the pressure difference between points at the crest and along the surface yields\be
\frac{1}{\rho} \, p(r) = \frac{1}{2} U_{\rm ref}^2 \left[ 1 - \left(\frac{r}{\ell_r}\right)^{2\alpha/(\pi-\alpha)}\right]. 
\ee
Only the horizontal velocity normal to the surface is expected to generate a pressure differential, i.e. the reference velocity $U_{\rm ref}$ would vanish if the surface does not present a component standing normal to the incoming flow direction. This effect can be taken into account by setting $U_{\rm ref} = U_k \hat{n}_x$, where 
\be \hat{n}_x = \frac{\partial h/\partial x}{|\boldsymbol{\nabla}_h h|}
\ee
is based on the horizontal gradient of the elevation map $h(x,y)$. For a surface that is inclined normal to the incoming flow, e.g. front faces of wall-attached cubes, $\hat{n}_x=1$ and thus $U_{\rm ref} = U_k$.  
On the side faces of such cubes, $\hat{n}_x$ and $U_{\rm ref}$ vanish, as the flow skims past such surfaces with no pressure buildup from the potential flow ramp model (see {Figure \ref{fig:sketchrampflow}}(\textit{c}).

Next, we compute the mean pressure on the ramp segment, $\overline{p}={\ell_r}^{-1}\int_{r=0}^{\ell_r}  p(r) \, dr$ which results in 
\be
\frac{1}{\rho} \,\, \overline{p}(x,y) =  (U_k \hat{n}_x)^2 \, \frac{\alpha(x,y)}{\pi+\alpha(x,y)}.  
\ee
To obtain the pressure force in the $x$ direction, i.e. in order to derive the form provided in Eq. \ref{eq:figeneral},  we multiply by the projected surface area vector  $|\boldsymbol{\nabla}_h h| \hat{n}_x \Delta x \Delta y$, where $\Delta x$ and $\Delta y$ are the horizontal spatial resolutions describing the surface (and the length $\ell_r$ used to evaluate the mean pressure is $\ell_r \sim (\Delta x \Delta y)^{1/2}$).
Including the shading function, the resulting local kinematic wall stress in the $x$-direction coming from pressure (i.e., dividing the pressure force by the planform area $\Delta x \Delta y$ and density) can be written as 
\be
\tau_{xz}^{p} = (U_k \, \hat n_x)^2 \, \frac{\alpha}{\pi+\alpha} \,\, \frac{\partial h}{\partial x}\, F^{\rm sh}.
\ee
For small slopes, we have $\alpha \approx |\boldsymbol{\nabla}_h h| << \pi$, and then the pressure drag is quadratic with slope (as is known to be the case for small-amplitude waves \cite{jeffreys1925formation}). However, for present applications in which for certain types of surfaces the local slopes could go up to   $\pi/2$ (vertical segments, although with finite resolution discretely representing $h(x,y)$, some small deviations from vertical are unavoidable), we do not use nor need this small angle approximation.   

Now, instead of following \cite{ayala2024moving}
in which only the windward facing portion of the surface feels a pressure force (the leeside portion was assumed to exhibit either flow separation or incipient separation such that the pressure force there was neglected), we here allow for pressure recovery for downward facing portions of the surface. To model absence of pressure recovery in separated or nearly separated region we here rely entirely on the sheltering function $F^{\rm sh}(x,y;\theta)$ treated in the subsection. Without flow separation, the potential flow ramp model predicts a flow along a downward ramp with the stagnation pressure at the bottom of the ramp and a resulting force in the negative $x$ direction, i.e. pressure recovery.  The sign of the resulting force is given by the surface slope in the $x$-direction ($\partial h/\partial x$) but the magnitude is the same independent of flow direction (for inviscid flow). By choosing to define  the angle $\alpha$ using an absolute value of the ramp slope and use $\partial h/\partial x$ to determine the direction of force, we enable pressure recovery on   backward sloping parts of the surface that are not in the sheltered portions of the surface (for surfaces to be studied in this work, however, this effect appears to be of negligible importance). 

\subsection{Wind-shade factor}

Combining the sheltering and inviscid pressure models, we write the total (kinematic) force for a flow in the $x$ direction ($i=1$)  as
\be
f_x = u_\tau^2 A =  U_k^2 \, \iint_A  \hat n_x^2 \,  \frac{\alpha}{\pi+\alpha} \, \frac{\partial h}{\partial x}  \, F^{\rm sh}(x,y;\theta) \, dx dy,
\ee
and the averaging is performed over the entire surface, i.e. over all $(x,y)$. This expression can be solved for the velocity $U_k$ normalized by friction velocity $u_\tau$ as follows:
\be
 U_k^+  = \frac{1}{\sqrt{ {\cal W}_{\rm L}} }, 
 \label{eq:ukvswl}
\ee
where it is useful to define the 
streamwise (longitudinal, ``L'') wind-shade factor ${\cal W}_{\rm L}$ using a surface average:
\be
{\cal W}_{\rm L} = \left< \hat n_x^2(x,y) \,  \frac{\alpha}{\pi+\alpha} \, 
\frac{\partial h}{\partial x}
 \,\, F^{\rm sh}(x,y;\theta)  \right>_{x,y}, ~~{\rm where} ~~ \alpha(x,y) =  \arctan|\boldsymbol{\nabla}_h h(x,y)| .
 \label{eq:defwindshade}
\ee 
It is important to note that, owing to the fact that potential flow is purely dependent on surface geometry, the wind-shade factor ${\cal W}_{\rm L}$ is also a purely geometric quantity depending only on the surface height distribution, the mean flow direction relative to the surface, and the assumed turbulence angle $\theta$.  

For future reference, we point out that for certain surfaces with directional preference (e.g. inclined ridge-like features) one can also define a transverse wind shade factor ${\cal W}_T$ according to
\be
{\cal W}_{\rm T} = \left< \hat n_x ^2 \,  \frac{\alpha}{\pi+\alpha} \, 
\frac{\partial h}{\partial y}
 \, F^{\rm sh}(x,y;\theta)  \right>_{x,y} .
\ee 
It involves the pressure force built up due to streamwise velocity but projected onto the transverse direction as a result of the local slope $\partial h/\partial y$.

\subsection{Reference height}

In order to relate the estimated drag force and wind shade factor to roughness ($z_0$) or equivalent sandgrain ($k_s$) roughness lengths, we must   choose an appropriate height  to evaluate the log law. The common wisdom is that the roughness sublayer extends to about 2-3 times the representative heights of roughness elements, \citep{Flack2007,Jimenez2004}, 
although further dependencies on in-plane roughness length-scales are also known to affect the height of the sublayer \citep{raupach1980wind,sharma2020turbulent,meyers2019decay,endrikat2022reorganisation}. 
For general surfaces, the height of roughness elements is difficult to specify and a more general definition for arbitrary functions $h(x,y)$ must be devised. Using the mean elevation $\overline{k} = \langle h \rangle$ as baseline height, we define a dominant positive height $k_{p}^\prime$ according to
\be
k_{p}^\prime = \langle [R(h^\prime)]^p\rangle^{1/p}, ~~{\rm where}~~
h^\prime = h-\langle h\rangle,  ~~~~ {\rm and} ~~ R(z)=z ~~ {\rm if} ~~ z \geq 0, ~~R(z)=0 ~~ {\rm if} ~~ z < 0,
\ee
is the ramp function. As $p \to \infty $, the scale $k_{p}^\prime$ tends to the maximum positive deviation above the mean height over the entire surface. A choice of $p=8$ turns out to be sufficiently high to both emphasize the highest points but still include weak contributions from the entire surface for statistical robustness, for practical applications. 

\begin{figure}
  \centerline{\includegraphics[width=14cm]{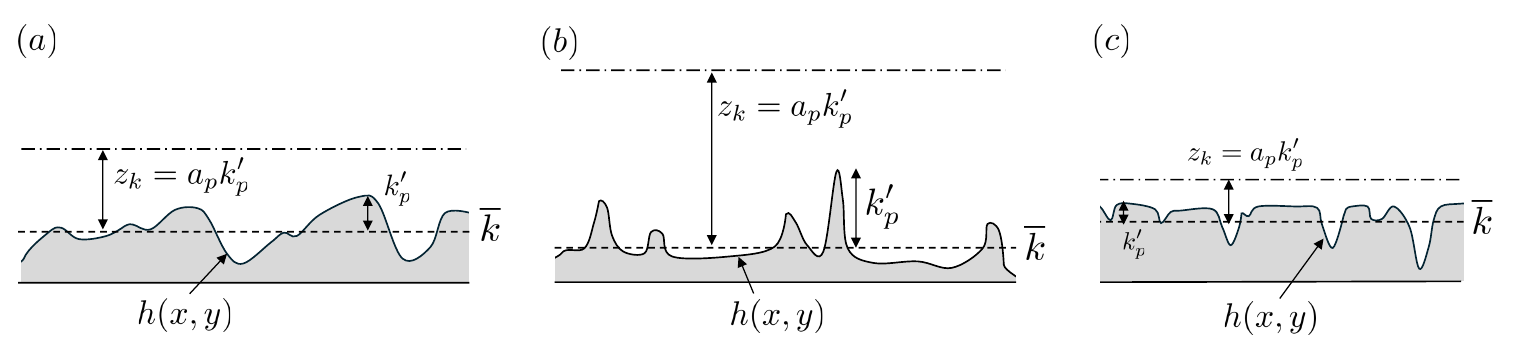}}
  \vspace{-0.1in}
  \caption{Sketch of surfaces with height distribution $h(x,y)$ and near zero (a), positive (b) and negative (c) skewness. Also shown are the mean height $\overline{k}$, the dominant positive height $k_{p}^\prime$, and the resulting reference height $a_p k_{p}^\prime$ (with $a_p \sim 3$ to be used in the model) above the mean height.}
   \vspace{-0.1in}
\label{fig:sketchsurf}
\end{figure}

As reference height where to evaluate the log-law and where we assume the reference velocity  $U_k$ is constant (i.e. the mean velocity only depends on $z$ and not on $(x,y)$), we select a height $\overline{k}+a_p k_{p}^\prime$, with $a_p=3$,
a choice motivated by the observations that the roughness sublayer extends to about  3 times the characteristic element heights. Since $a_p$ is somewhat arbitrarily chosen we can regard this parameter as an adjustable one, but as a first approximation the choice $a_p=3$ appears reasonable. The sketch in figure \ref{fig:sketchsurf} illustrates three types of surfaces and the resulting reference heights given the maximal positive height fluctuation away from the mean. 

With these assumptions,  equivalent roughness scales $z_{0}$ and sandgrain roughness heights $k_s$ can be obtained as usual \citep{Jimenez2004} 
from  
 $ U_k^+ =  (1/\kappa) \ln(z_k/z_0) 
 = (1/\kappa) \ln(z_k/k_s) + 8.5 $,
with $z_k=a_p k_{p}^\prime$, and hence 
\be
z_{0} = a_p \, k_{p}^\prime \, \exp\left(-\kappa  \, U_k^+ \right),
\label{eq:z0form}
\ee
and similarly for $k_s$.  Additionally, an overall reference scale for surface height will be chosen as $k_{\rm rms} = \langle (h-\overline{k})^2\rangle^{1/2}$,  the root-mean-square of the surface height function $h(x,y)$ over all $(x,y)$. This choice facilitates comparison with datasets for which $k_{\rm rms}$ is often prescribed and known.  

Finally the wind-shade model for sandgrain-roughness normalized by rms height is given by

\be
\frac{k_{\rm s-mod}}{k_{\rm rms}} = a_p \, \frac{k_p^\prime}{k_{\rm rms}} \, \exp\left[-\kappa \, (U_k^+ - 8.5) \right],
\label{eq:ksmod}
\ee
with $U_k^+ = {\cal W}_L^{-1/2}$ for the case of fully rough conditions (see \S\ref{sec:transitional} below for transitionally rough cases).

\subsection{Turbulence spreading angle}
The angle $\theta$ describing the effect of turbulent mixing at the reference height determines the strength of the sheltering effect. If the angle is large, the sheltering region will decrease rapidly in size and increase drag due to larger exposed downstream roughness features (see figure \ref{fig:sampleshaded}). Conversely, for smoothly rounded surfaces, increasing $\theta$ will lead to delayed separation on the back sides of rounded roughness elements and hence to partial pressure recovery which would decrease the drag. 

For the default model, we assume a constant angle and chose 10 degrees as representative of values expected from literature \citep[e.g.][]{rowin2024modelling}.  Another option, as proposed in \cite{yang2016exponential}, is to assume that the turbulence spreading angle $\theta$ is proportional the ratio of vertical root-mean-square turbulence velocity (which is of same order as $u_\tau$) and the streamwise advection velocity $U_k$ at the reference height, i.e, 
\be 
\tan \theta = \frac{u_\tau}{U_k}=\frac{1}{U_k^+} = {\cal W}_{\rm L}^{1/2}.
\label{eq:thetavsWL}
\ee
The wind shade factor and the angle must then be determined iteratively. 
(Eq. \ref{eq:thetavsWL} is meant for cases  when ${\cal W}_T=0$ since otherwise the friction velocity $u_\tau$ would also be expected to depend on ${\cal W}_T$).

\subsection{Near wall velocity profile}
A further refinement relates to the assumed velocity near the roughness elements. The definition of the wind-shade factor ${\cal W}_{\rm L}$ relies on the assumption that the velocity is constant (plug flow) from $z_k$ down to near the crest of any of the surface features or roughness elements. For surfaces with a few large-scale elements and many very small-scale elements (e.g. multiscale roughness), this may lead to a drag over-prediction caused by the small-scale elements that in reality might be exposed to a smaller velocity there. A possible remedy is to assume a 1/7 velocity power-law profile impinging on the roughness elements of the form $U(z) = U_k [(h(x,y)-h_{\rm min})/(h_{\rm max}-h_{\rm min})]^{1/7}$. In this case, the  definition of the wind-shade factor becomes

\be
{\cal W}_{\rm L,vel} = \left< \left(\frac{h(x,y)-h_{\rm min}}{h_{\rm max}-h_{\rm min}}\right)^{2/7}\, \hat n_x^2(x,y) \,  \frac{\alpha}{\pi+\alpha} \, 
\frac{\partial h}{\partial x}
 \,\, F^{\rm sh}(x,y;\theta)  \right>_{x,y}.
 \label{eq:defwindshadevel}
\ee 
The introduction of a factor that decreases to zero near the minimum of the surface may improve the predictive power of the wind-shade factor. Note that this expression is still purely dependent on the surface geometry, although as with the pressure distribution model, it is fluid-mechanics inspired. 

At this stage, it is useful also to comment on connections to existing parameters. For instance,  for vertical surfaces with $\alpha=\pi/2$, the slope correction term $\alpha/(\pi + \alpha) = 1/3$, and the flow-direction correction is $\hat{n}_x^2 = 1$, and if the wind shading $F^\text{sh} = 1$ if $\partial h/\partial x > 0$ and $0$ otherwise, then we recover one third of the frontal solidity as the model's prediction. In terms of the formula of \citet{Macdonald1998} that accounts for densely packed roughness features, choosing as wind shading $F^\text{sh} = 1$ for sparse roughness and $F^\text{sh} = 0$ for dense roughness is similar to the damping factor introduced that is also $1$ and $0$ as the plan solidity varies between $0$ and $1$ respectively. However, in the present model $F^\text{sh}$ is linked in more detail with the underlying flow physics given its detailed $(x,y)$ dependence that can also be locally correlated to the value of $\alpha(x,y)$.
The inclusion of the slope correction $\alpha/(\pi+\alpha)$ is expected to discriminate between spiky versus undulating surfaces, an argument usually given to include the surface skewness as relevant parameter. Moreover, directional and clustering effects are all naturally accounted for.

\section{Effects of viscosity: transitionally rough flows}
\label{sec:transitional}
Since in many cases viscous effects are expected to be important (transitionally rough cases), we now include  the contributions from viscous stresses. Following \citet{raupach1992drag}, we model the local kinematic viscous stress as a function of the assumed velocity $U_k$ using a friction factor 
\be
 {\tau^{\nu}_{xz}} = \frac{1}{2} \, c_f(Re_k) \, U^2_k.
\ee
The smooth-surface friction factor $c_f$ can be written in terms of a Reynolds number $Re_k=z_k U_k/\nu$ based on the height $z_k=a_p k_p^\prime$ and velocity there, $U_k$. The viscous stress  {$\tau^{\nu}_{xz}$} acts in the the unsheltered regions. To include the sheltering we multiply by the average of the sheltering function $\langle F^{\rm sh}\rangle$. However, for transitionally and hydrodynamically smooth surfaces the sheltering effect decreases and must entirely vanish in the limit of a smooth surface, regardless of the existence of geometric roughness. Hence, we introduce the average viscous sheltering factor 
\be
\overline{F}_\nu^{\rm sh}(z_k^+) = \langle F^{\rm sh}\rangle + 
(1- \langle F^{\rm sh}\rangle ) \, \left[1+\left(  {z_k^+}/{10}\right)^2 \right]^{-1/2},
\ee
which switches from the rough limit $\overline{F}_\nu^{\rm sh} = \langle F^{\rm sh}\rangle$ (expected when $z_k^+ >> 10$) to $\overline{F}_\nu^{\rm sh} =1$ when approaching hydrodynamically smooth surface behavior (expected when $z_k^+ << 10$). Strictly speaking, we should make a similar adjustment to the wind shade factor ${\cal W}_{\rm L}$, since in the limit of transitionally and hydrodynamically smooth surfaces one would expect an absence of flow separation, and therefore full pressure recovery, over the backward facing surfaces. However, one would then need to transition to a low-Reynolds number model for pressure, instead of the quadratic inviscid model used for defining ${\cal W}_{\rm L}$. In practice, since the viscous $c_f$ term dominates at low $Re_k$, such an adjustment would have very little impact on results, and in the interests of simplicity such adjustments are omitted, and the definition of ${\cal W}_{\rm L}$  based on a purely inviscid quadratic drag law is retained.

The total  drag force can then be written as  
\be
f_x = u_\tau^2 A = U_k^2 \,\, {\cal W}_{\rm L} \, A  
+ U_k^2 \,   \frac{1}{2} \, c_f(Re_k) \,\overline{F}_\nu^{\rm sh}(z_k^+) \, A,
\ee
or, solving again for $U_k^+$
\be
 U_k^+ =  \left[{\cal W}_{\rm L} +  \frac{1}{2}  \, c_f(Re_k) \overline{F}_\nu^{\rm sh}(z_k^+)   \right]^{-1/2}
\label{eq:mombalance}
\ee
since $U_k$ (and thus $Re_k$) are assumed to be the same over entire surface. We also recall that $Re_k = U_k z_k/\nu = U_k^+ \, z_k^+ = U_k^+\, (z_k /k_{\rm rms}) \, k_{\rm rms}^+$ with $z_k=a_p k_p^\prime$, which for a prescribed value of $k_{\rm rms}^+$ and known geometric ratio $(a_p k_p^\prime/k_{\rm rms})$ enables us to determine the Reynolds number in terms of $U_k^+$. 

Next, the friction factor must be determined. We use the generalized Moody diagram fit \citep{meneveau2020note}, appropriately rewritten in its simplest form according to   
\be
c_f(Re_k) = 0.0288 \, Re_k^{-1/5} \, \left(1+577 Re_k^{-6/5}\right)^{2/3}.
\ee
This is a fit to the numerical solution of the 1D equilibrium fully developed (parallel flow) wall-bounded flow problem. It represents a generalization of the Moody diagram method relating wall stress to bulk velocity but here used only for the smooth line of the Moody diagram, or the skin-friction law for a smooth wall, where $\text{Re}_k$ replaces the outer Reynolds number. Also the input parameter $\text{Re}_k$ is expressed at some reference height $z_k^+$ that may fall in the logarithmic or viscous sublayer, instead of an outer-layer height as is usual for the Moody diagram. For more details, see \cite{meneveau2020note}. 
Thus in the limit of small Reynolds number (i.e. when $z_k^+ \sim 1$, and the velocity profile up to maximal height is linear), the limiting behavior is $c_f \to 2/Re_k$ and $\overline{F}_\nu^{\rm sh}(z_k^+)\to 1$. For $k \sim x^{1/2}$, this trend is similar to Blasius. For increasing Reynolds number, the friction factor fitting function goes to a $Re_k^{1/5}$ high Reynolds number behavior. 

To solve Eq. \ref{eq:mombalance} we can employ numerical iteration. Since the weakest dependence is on the $c_f$ function, it is convenient to simply iterate the rapidly converging expression
\be
{U_k^+}^{(n+1)} =   \left(  {\cal W}_{\rm L} +  \frac{1}{2}  c_f( \, {U_k^+}^{(n)}  \, z_k^+ ) \,\, \overline{F}_\nu^{\rm sh}   \right)^{-1/2},
\label{eq:ukplusiter}
\ee
 always recalling that $z_k^+ = (a_p k_p^\prime/k_{\rm rms}) \, k_{\rm rms}^+$, and starting with some initial guess (e.g. near frictionless  $c_f^{(n=0)}=10^{-14}$). Once $U_k^+$ has been determined, the equivalent sandgrain roughness $k_s$ is again determined via Eq. \ref{eq:ksmod}. 

\section{Results}\label{sec:results}

\subsection{Suite of surfaces from roughness database and others}

\begin{figure}
    \centering
    \begin{subfigure}{0.3\textwidth}
        \includegraphics[width=1.1\linewidth]{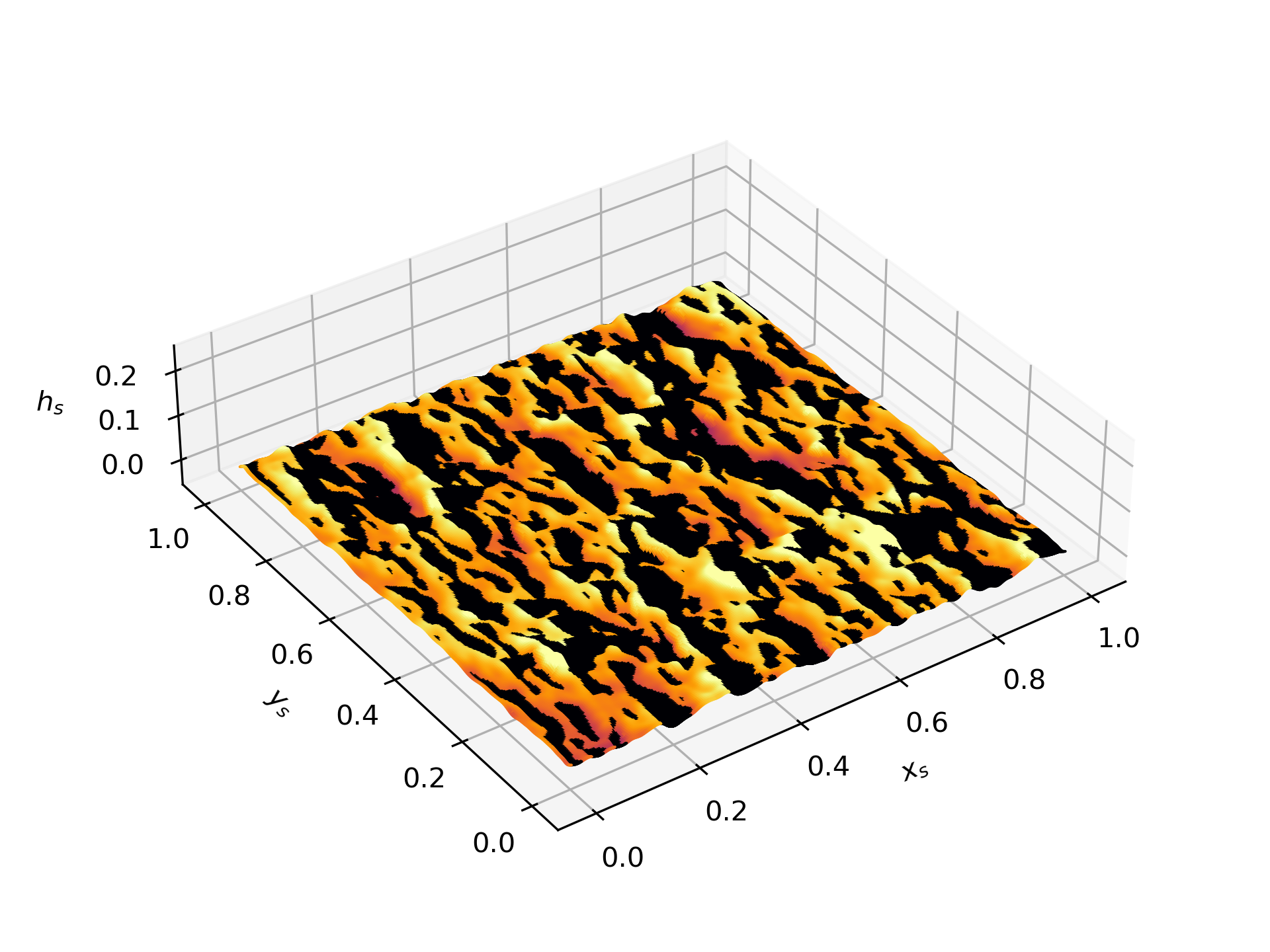}
        \vspace{-0.3in}
        \caption{ }
    \end{subfigure} \hfill
    \begin{subfigure}{0.3\textwidth}
        \includegraphics[width=1.1\linewidth]{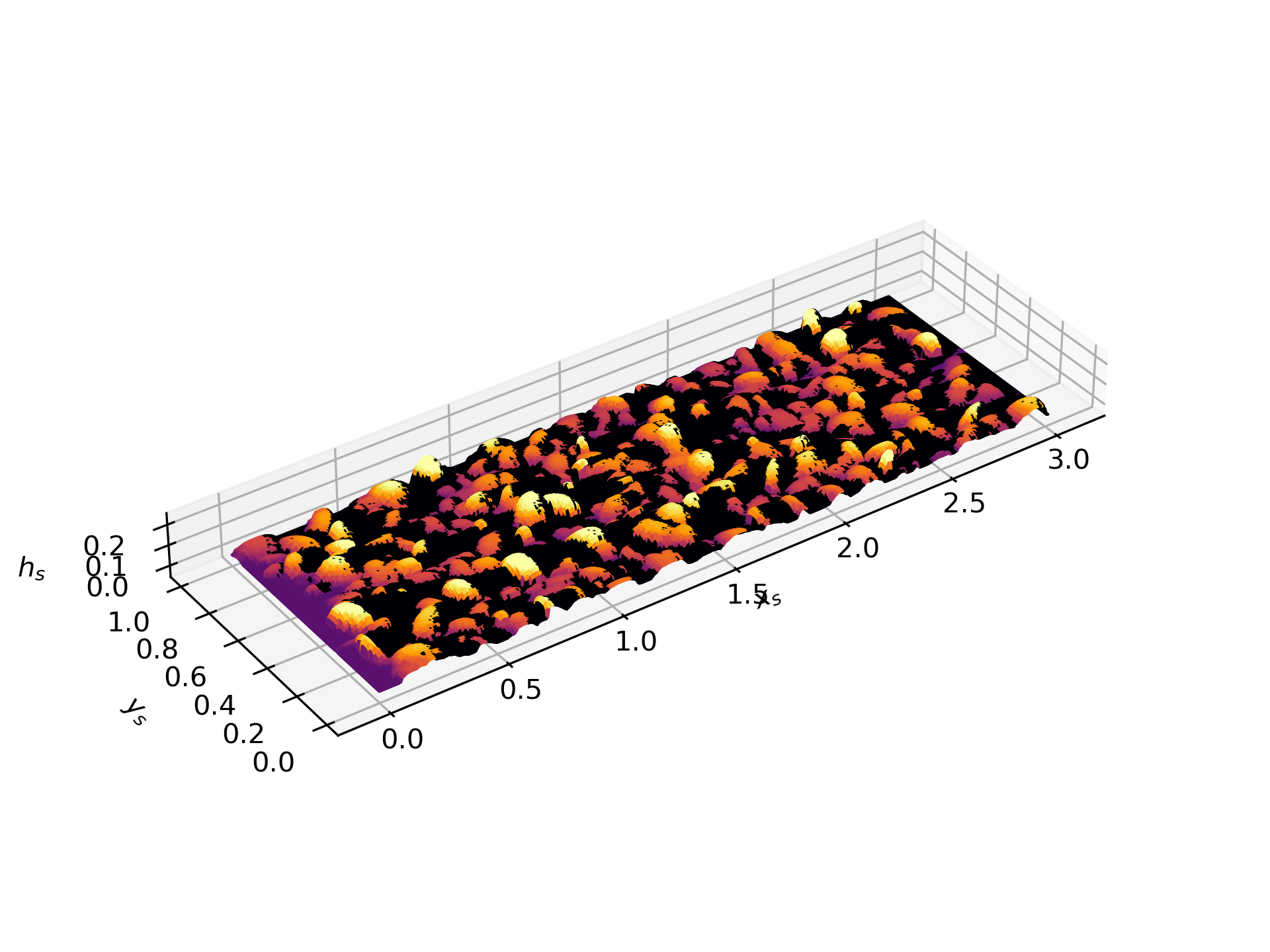}
             \vspace{-0.3in}
        \caption{ }
    \end{subfigure}\hfill
        \begin{subfigure}{0.3\textwidth}
        \includegraphics[width=1.1\linewidth]{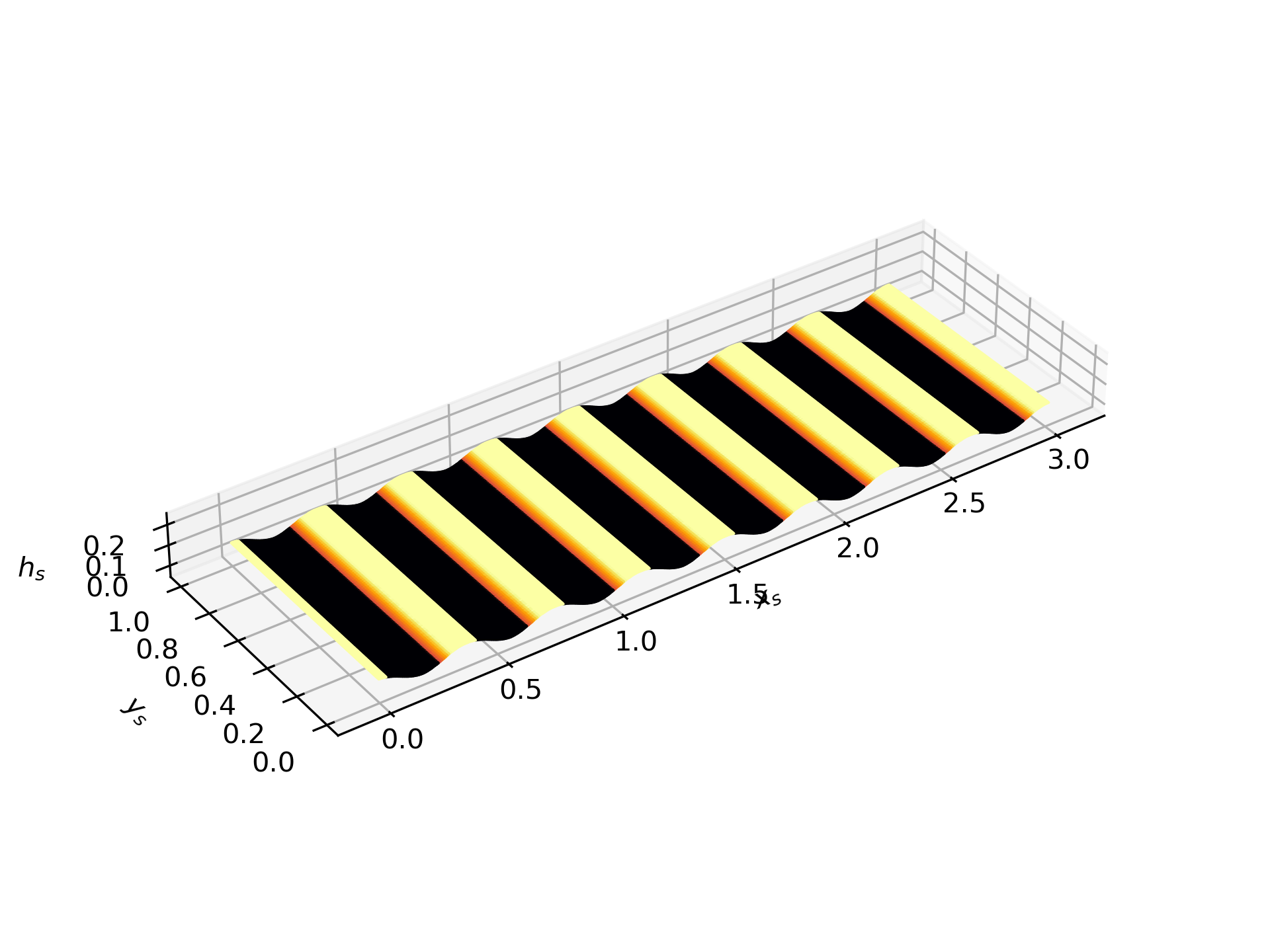}
             \vspace{-0.3in}
        \caption{ }
    \end{subfigure}\hfill
\vspace{-0.0in}
    \begin{subfigure}{0.3\textwidth}
        \includegraphics[width=1.1\linewidth]{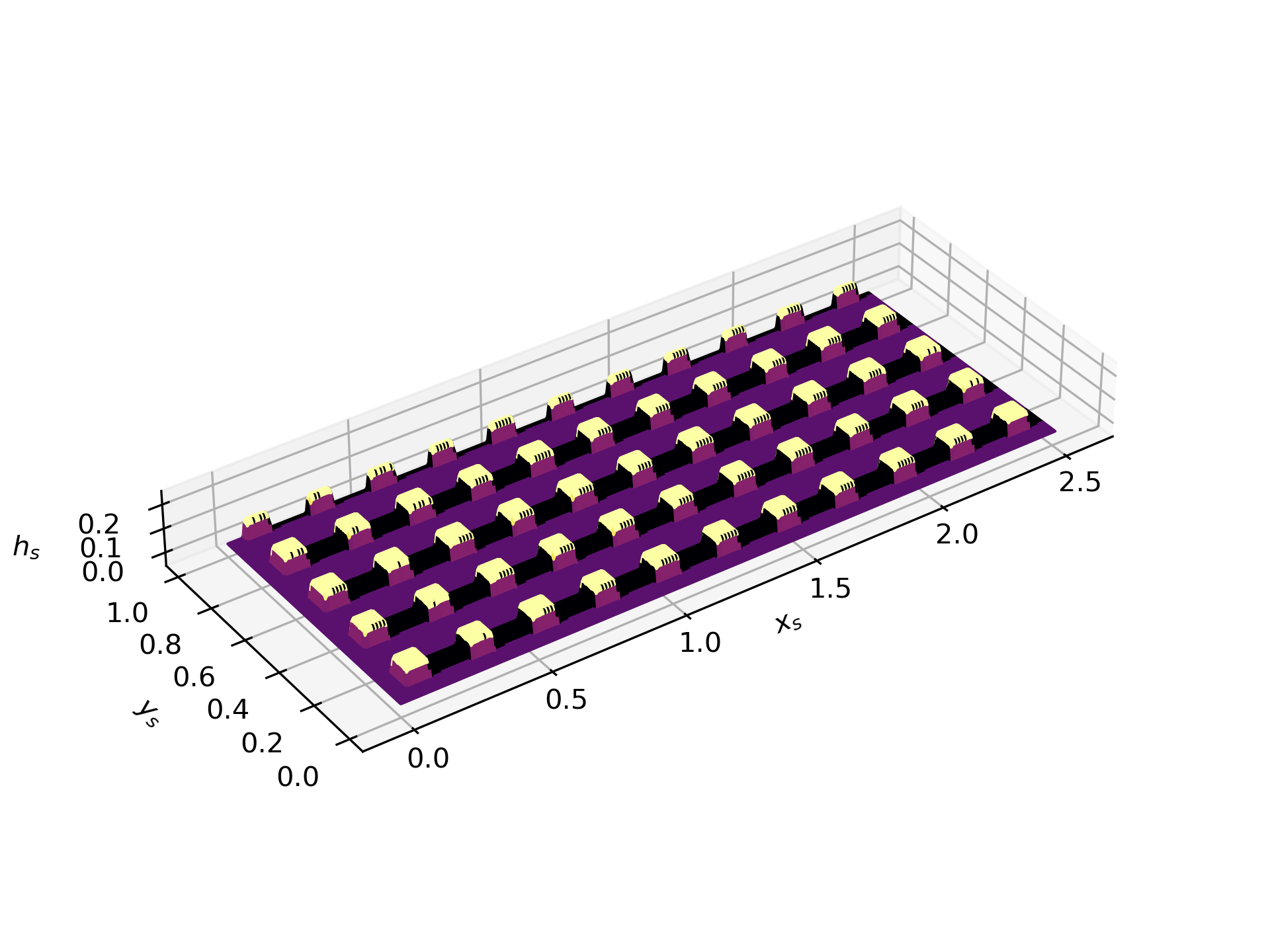}
             \vspace{-0.4in}
        \caption{ }
    \end{subfigure} \hfill
    \begin{subfigure}{0.3\textwidth}
        \includegraphics[width=1.1\linewidth]{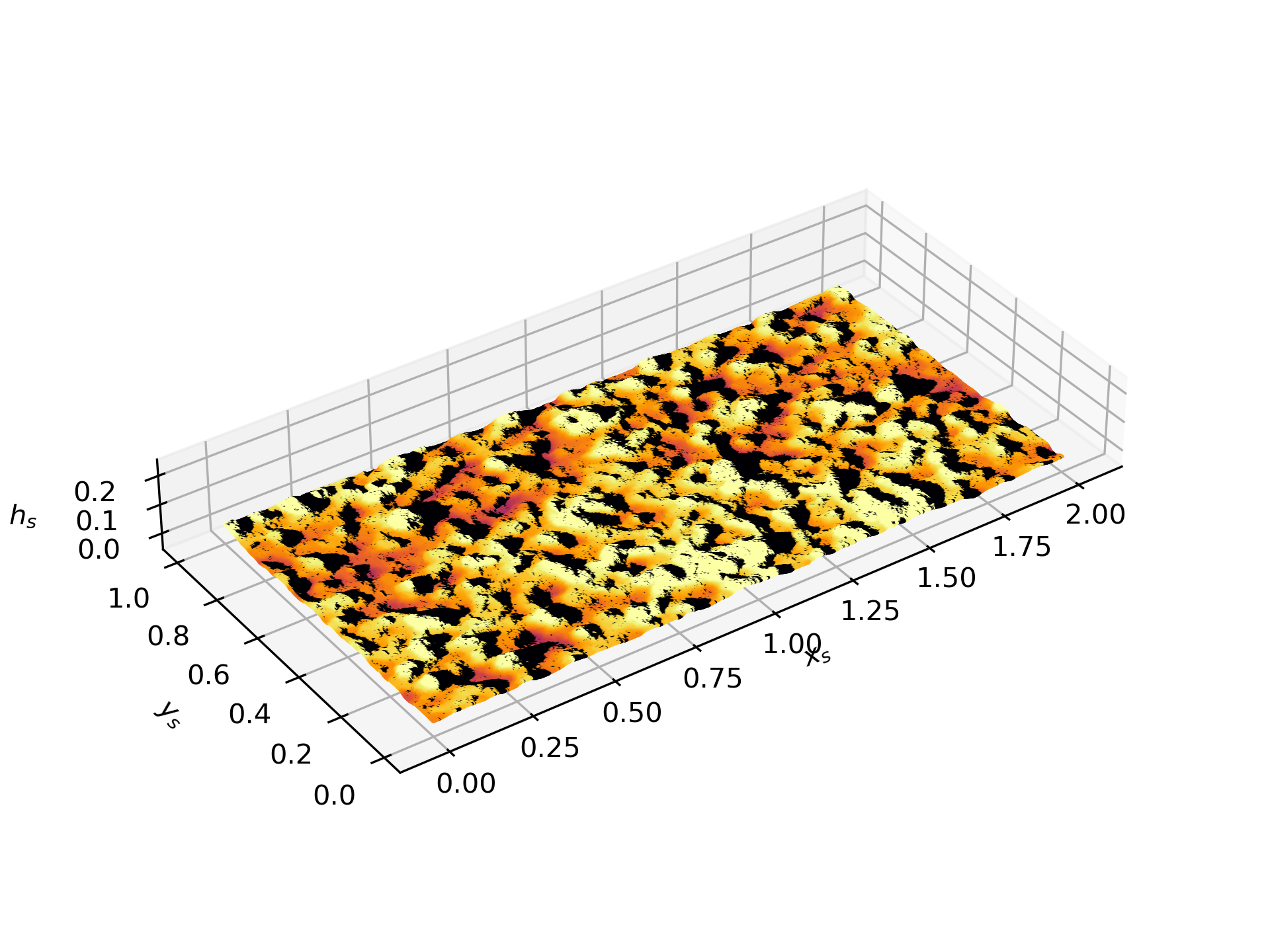}
             \vspace{-0.4in}
        \caption{ }
    \end{subfigure}\hfill
        \begin{subfigure}{0.3\textwidth}
        \includegraphics[width=1.1\linewidth]{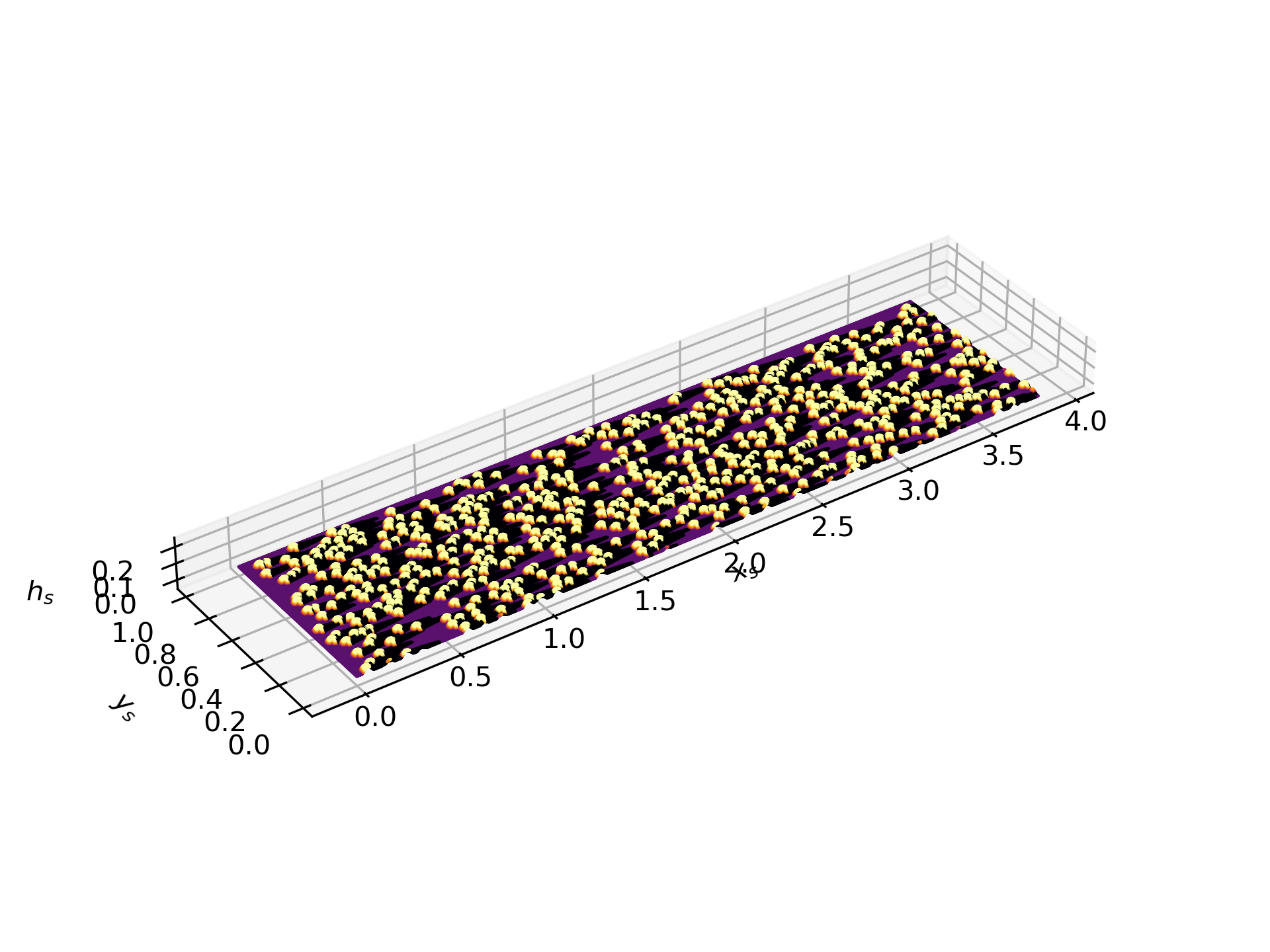}
             \vspace{-0.4in}
        \caption{ }
    \end{subfigure}\hfill
\vspace{-0.0in}
    \begin{subfigure}{0.3\textwidth}
        \includegraphics[width=1.1\linewidth]{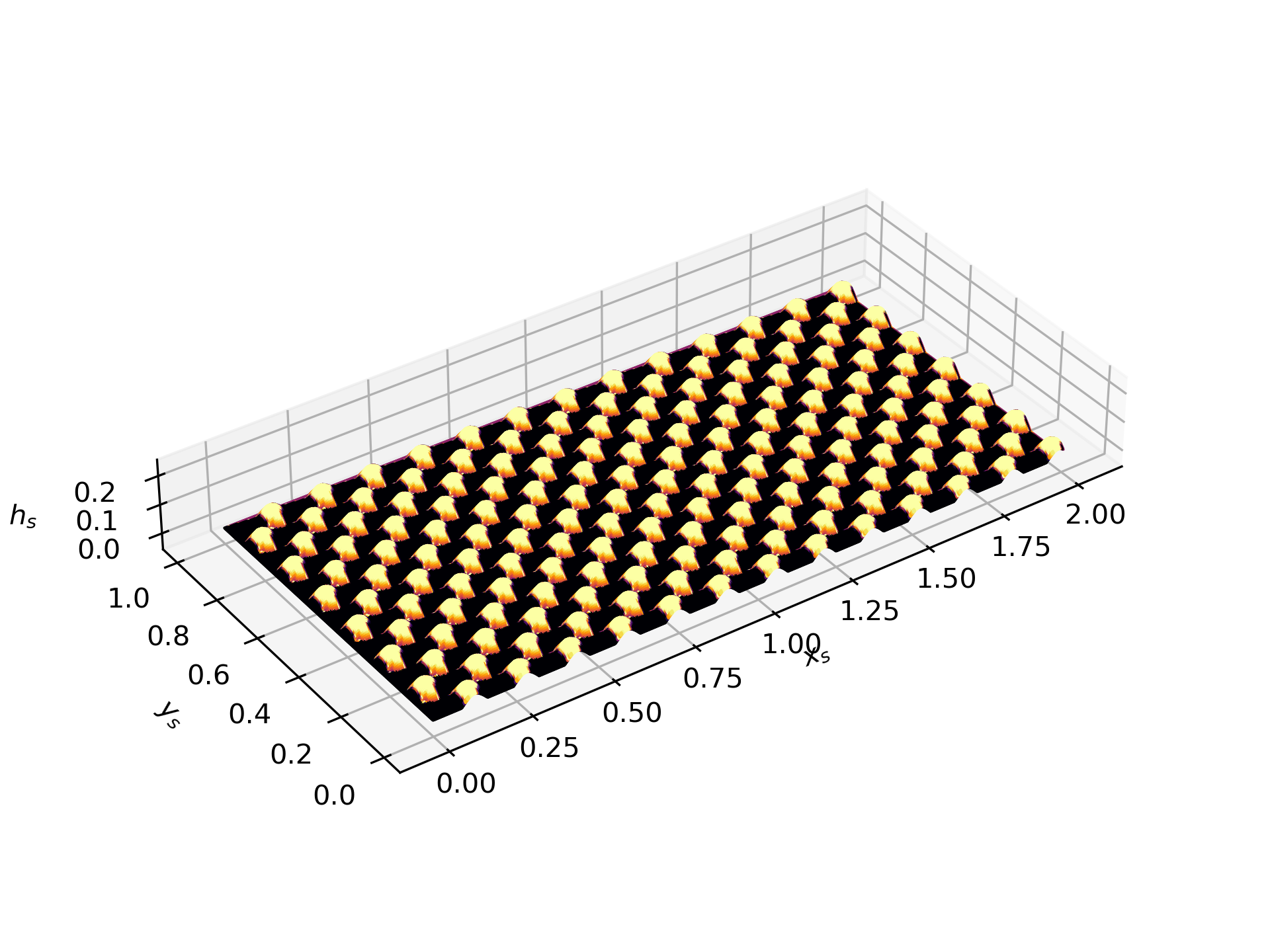}
             \vspace{-0.4in}
        \caption{ }
    \end{subfigure} \hfill
    \begin{subfigure}{0.3\textwidth}
        \includegraphics[width=1.1\linewidth]{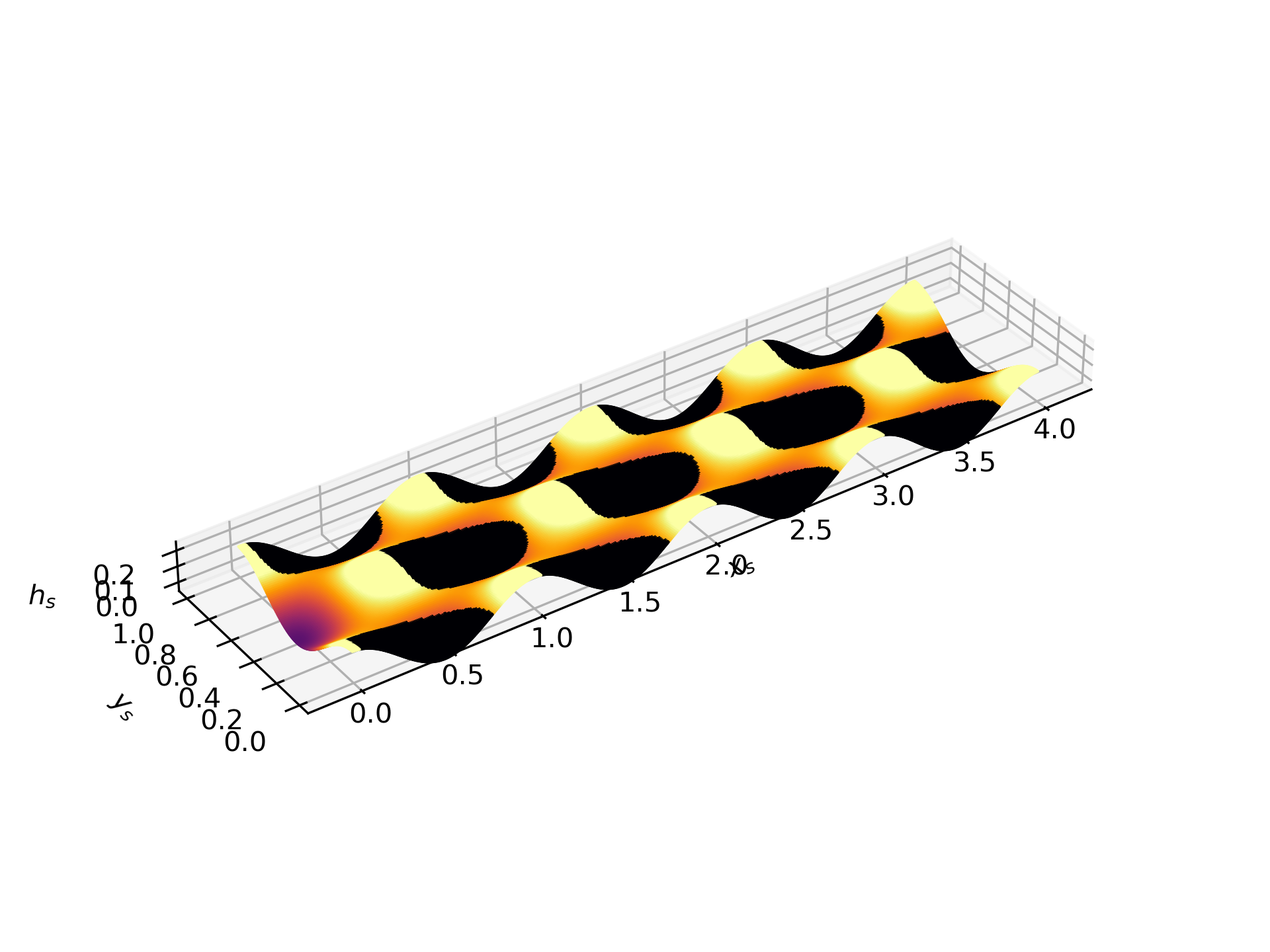}
             \vspace{-0.4in}
        \caption{ }
    \end{subfigure}\hfill
        \begin{subfigure}{0.3\textwidth}
        \includegraphics[width=1.1\linewidth]{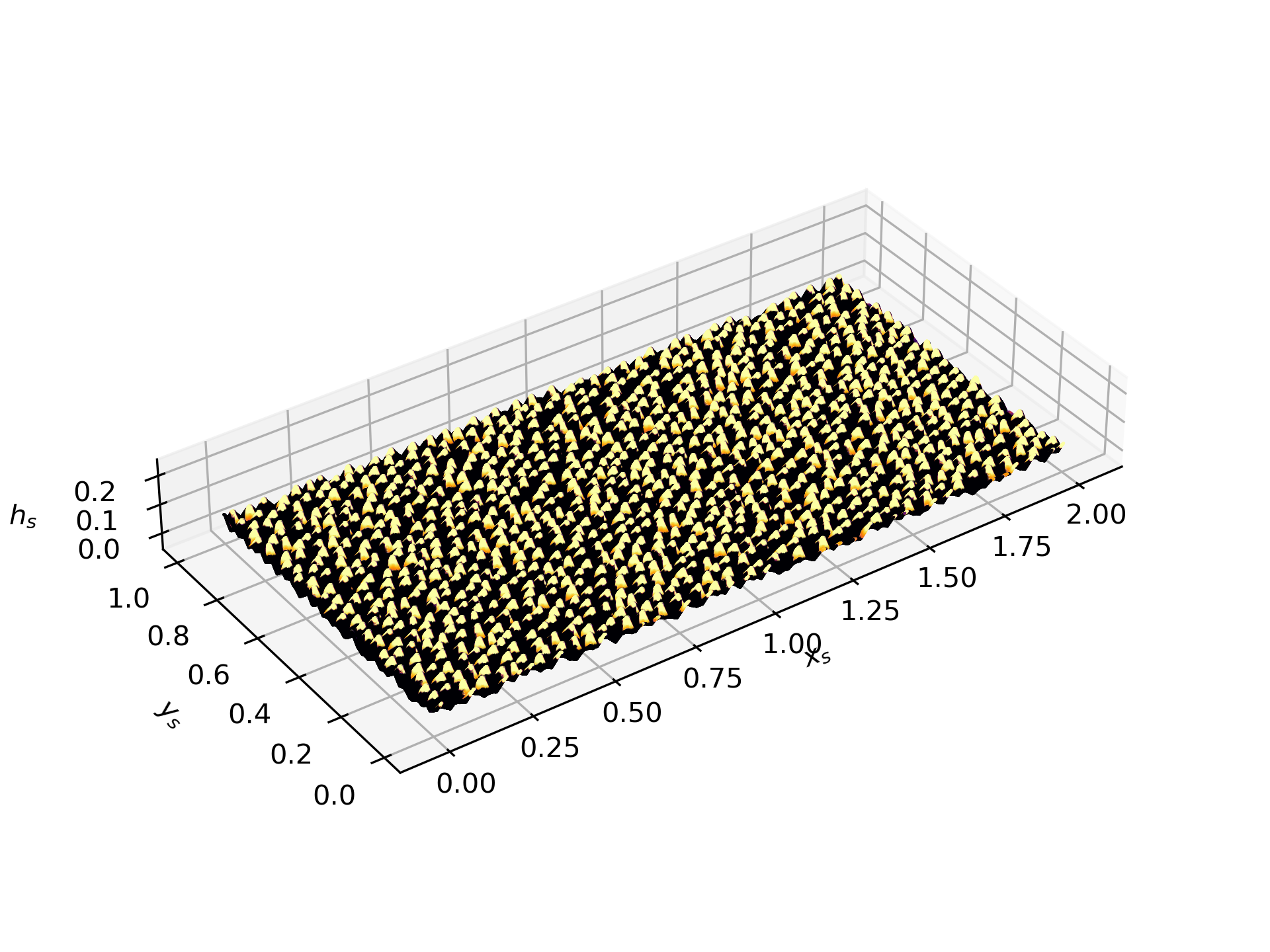}
             \vspace{-0.4in}
        \caption{ }
    \end{subfigure}\hfill
  \vspace{-0.0in}
    \begin{subfigure}{0.3\textwidth}
        \includegraphics[width=1.1\linewidth]{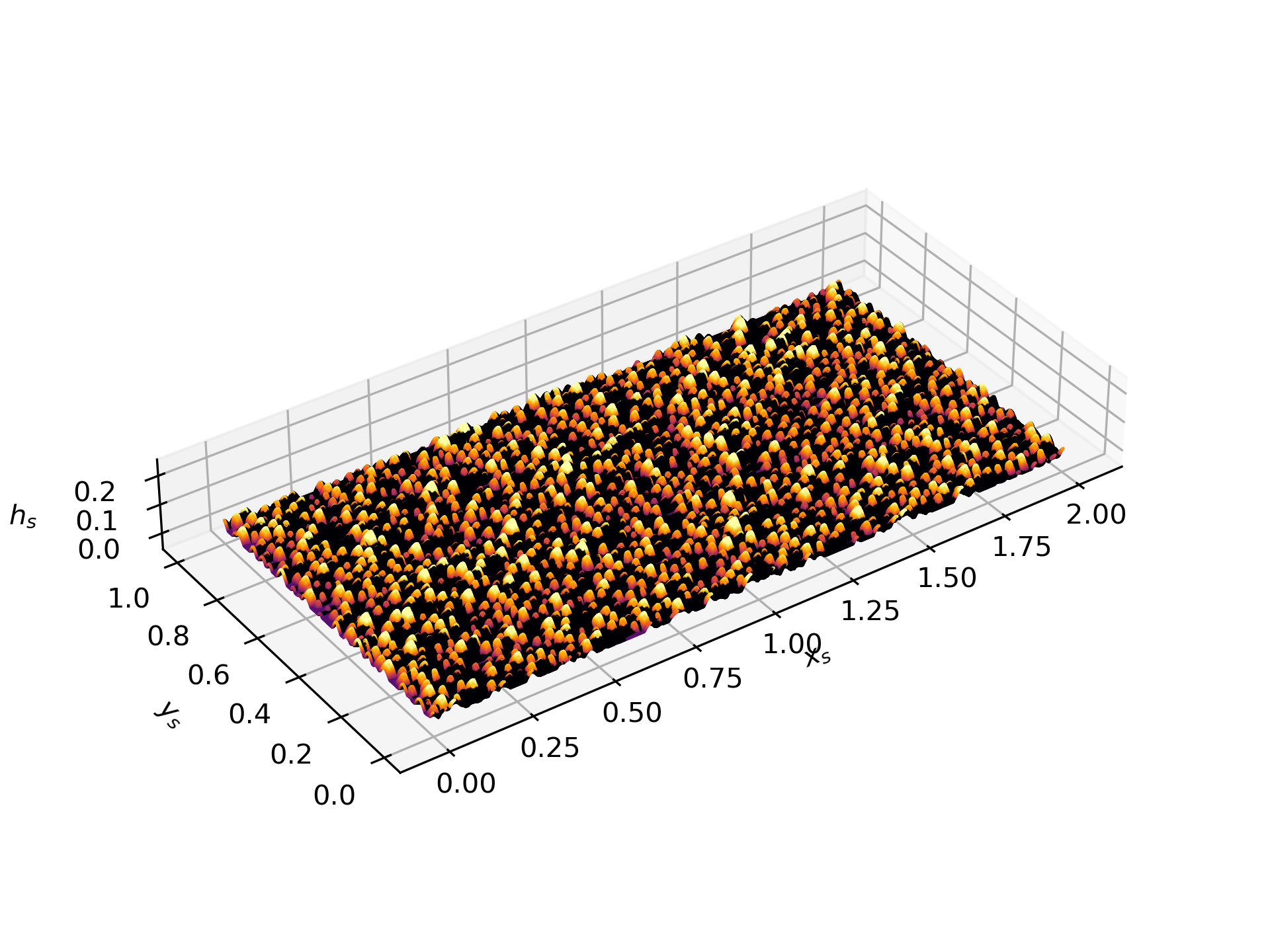}
             \vspace{-0.3in}
        \caption{ }
    \end{subfigure} \hfill
    \begin{subfigure}{0.3\textwidth}
        \includegraphics[width=1.1\linewidth]{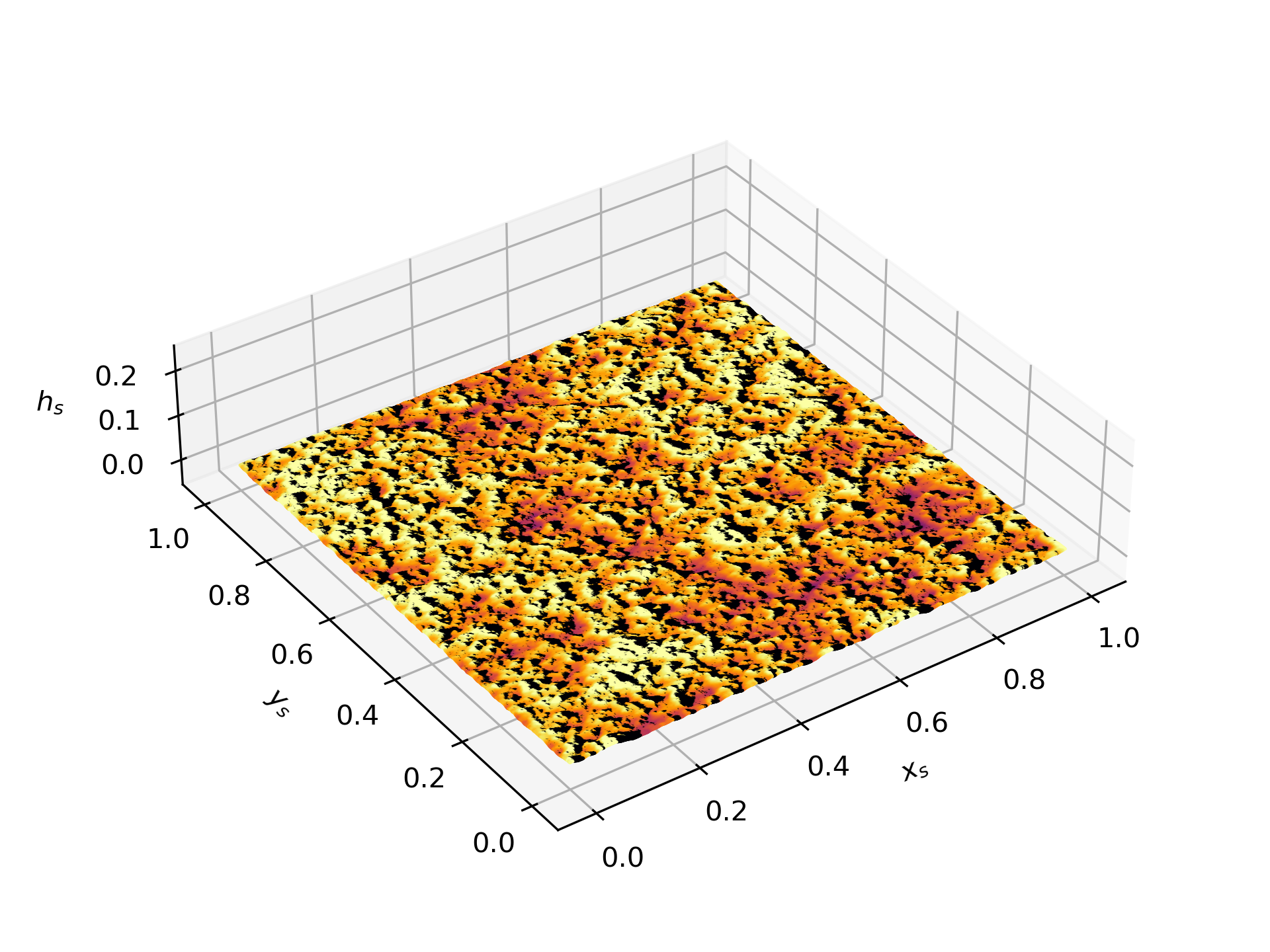}
             \vspace{-0.3in}
        \caption{ }
    \end{subfigure}\hfill
        \begin{subfigure}{0.3\textwidth}
        \includegraphics[width=1.1\linewidth]{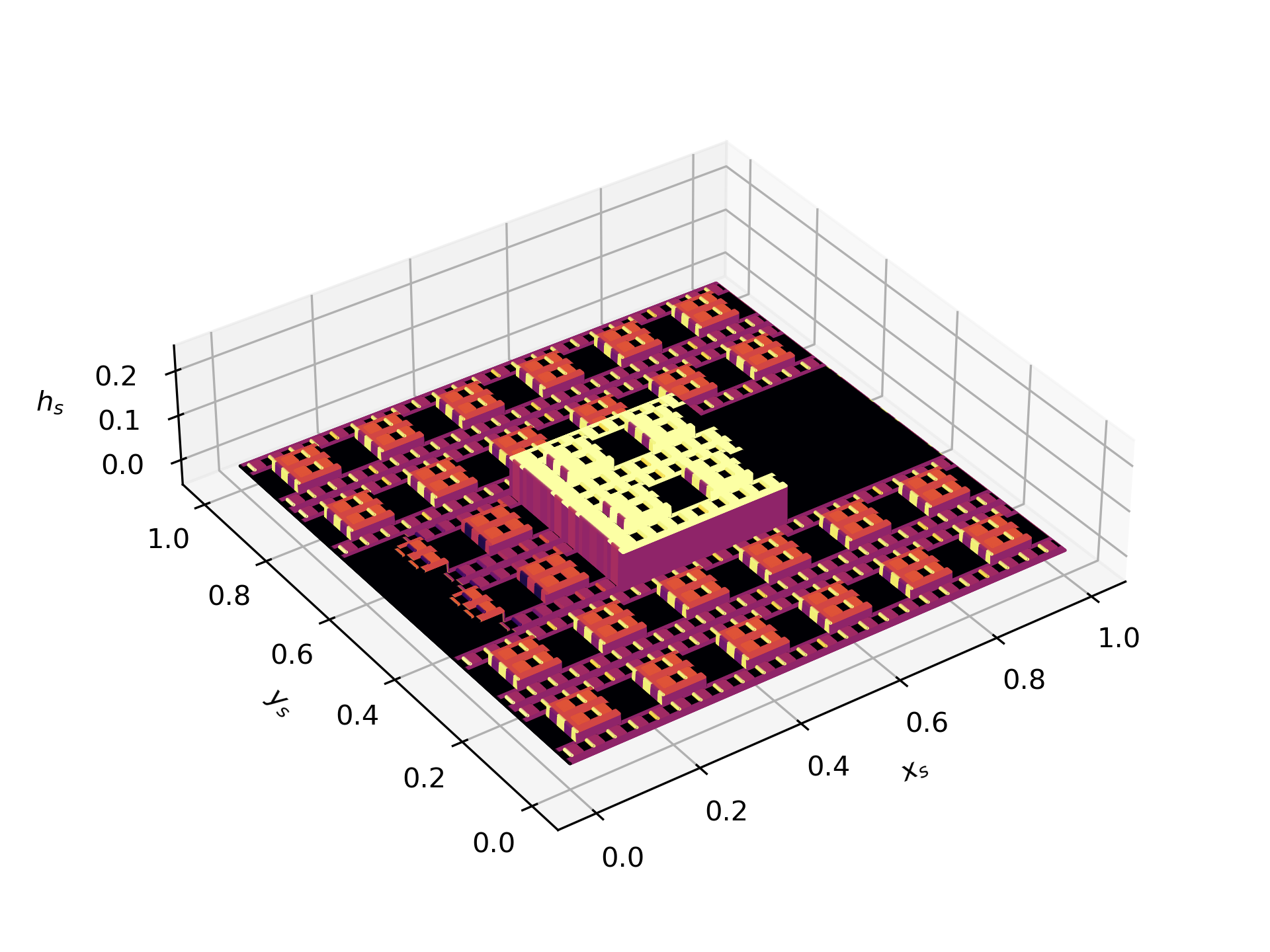}
             \vspace{-0.3in}
        \caption{ }
    \end{subfigure}\hfill
\vspace{-0.0in}
\caption{Representative sample of 12 (out of 104) surfaces considered in this study. Black regions are sheltered regions for $\theta=10^o$. Details about the surfaces are provided in appendix A. Surfaces shown are from (a): \cite{jelly2022impact}, $y$-aligned ridges,
(b): from \cite{jouybari2021data}: case C19 (random ellipsoids), 
(c): case C29 (sinusoidal), (d) and case C45 (wall-attached cubes),
(e): from \cite{flack2020skin}: rough surface case 1 (panel 1),
(f): from \cite{womack2022turbulent} truncated cones, random case R48,
and (g): regular staggered case S57,
(h): from \cite{rowin2024modelling} intermediate eggbox (case 0p018),
(i): from \cite{forooghi2017toward} case A7060, and 
(j): case C7088, (k): from 
\cite{barros2018measurements} power-law random surface with spectral exponent p=-0.5 , and (l) 3-generation multiscale block surface by 
\cite{medjnoun2021turbulent}, case iter 123. The surface height is indicated by colors, ranging from light yellow to dark purple from highest to lowest elevation of each surface, respectively. Shaded portions of the surface are indicated in black. 
}
 \label{fig:allfigs}
\end{figure}

In this section we apply the wind-shade roughness model to over 100 surfaces for which the full height distribution $h(x,y)$ is known and the equivalent sandgrain roughness $k_s$ has been measured. Appendix A lists the datasets considered (most of the surfaces are obtained from a public database 
 at \href{https://roughnessdatabase.org}{\color{blue}https://roughnessdatabase.org}). Summarizing, 
they include 3 rough turbine blade-type surfaces  from \cite{jelly2022impact} (isotropic and with dominant streamwise or spanwise aligned ridges), 26 cases studied numerically by \cite{jouybari2021data} including several types of {sandgrain} bumps, regular bumps, sinusoidal surfaces, and one wall-attached cube case, 7 cases of rough surfaces studied experimentally in \cite{flack2020skin}, truncated cones in random arrangements and  along regular staggered arrays, each in 8 different densities studied experimentally in  \cite{womack2022turbulent}. Also incuded are 3 eggbox sinusoidal surfaces from the numerical study of \cite{rowin2024modelling}, a large collection of 31 surfaces from \cite{forooghi2017toward} studied numerically,  3 power-law surfaces with spectral exponents -1.5, -1 and -0.5 studied experimentally by \cite{barros2018measurements}, surfaces with   closely packed cubes with 7 different densities from \cite{xu2021flow} and, finally, 4 cases of multiscale  blocks surfaces studied experimentally by \cite{medjnoun2021turbulent} with up to 3 generations. For a subset of 12 of these surfaces, the panels in Fig. \ref{fig:allfigs} show the surface elevation and sheltered regions shown as black shadows.

\begin{figure}
\centering
    \begin{subfigure}{0.45\textwidth}
    \hspace{-0.2in}
        \includegraphics[width=1.3\linewidth]{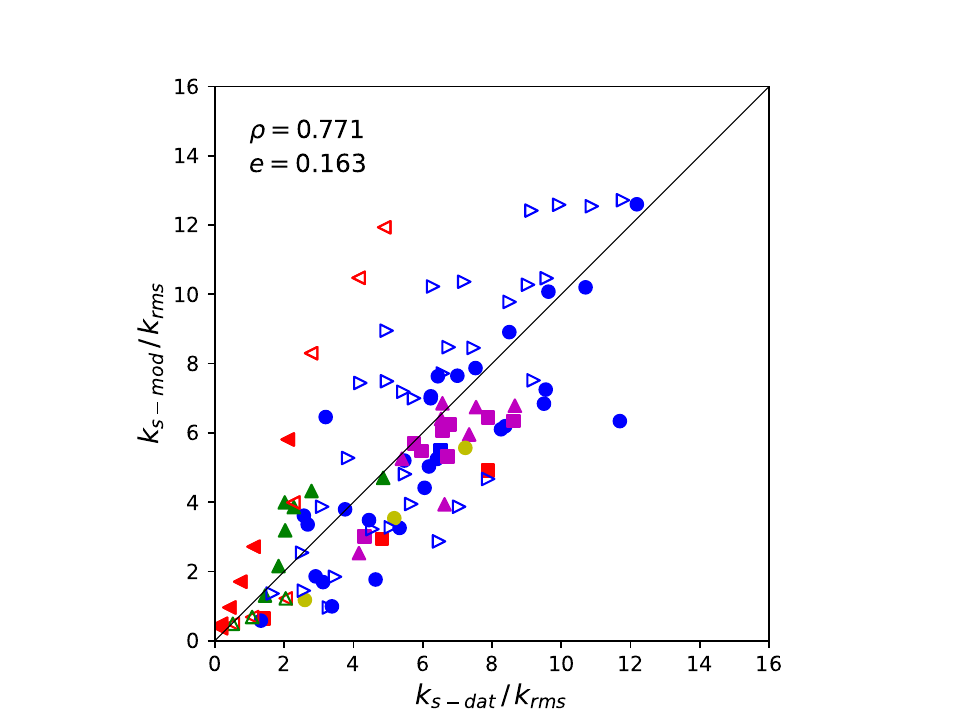}
        \caption{ }
    \end{subfigure} \hfill
     \begin{subfigure}{0.45\textwidth}
          \hspace{-0.1in}
        \includegraphics[width=1.1\linewidth]{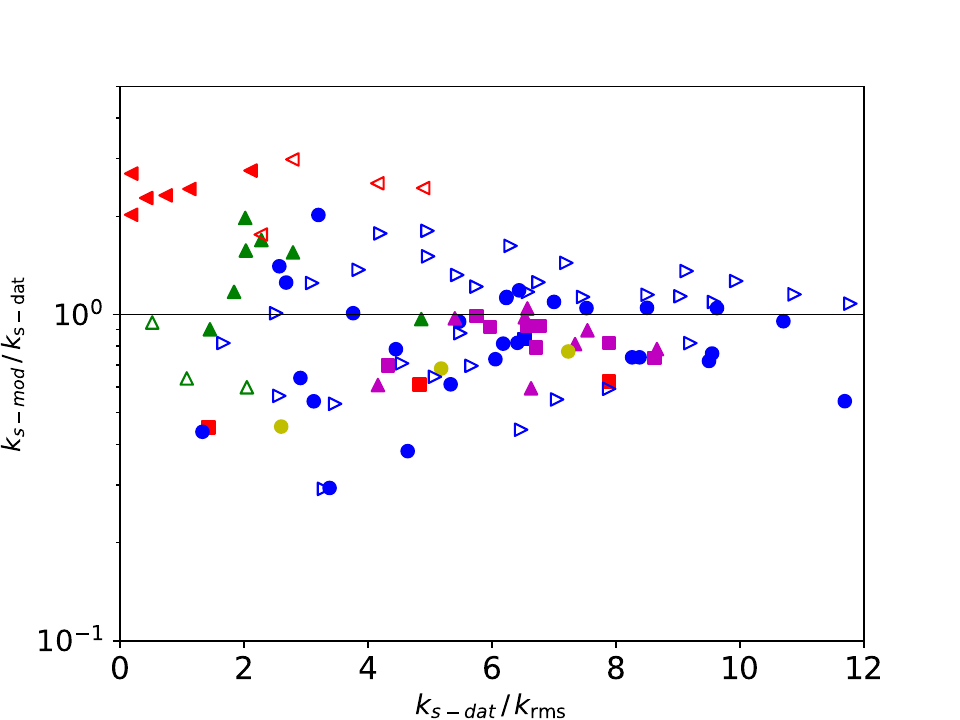}
        \caption{ }
    \end{subfigure}  
   \caption{Sandgrain roughness predicted by the model with $\theta=10$ degrees and $a_p=3$, versus measured $k_s$ values from data, for all 104 surface cases considered in this work. Data from \cite{jelly2022impact} (red solid squares), 26 cases from \cite{jouybari2021data} (blue solid circles), wall attached cubes from \cite{jouybari2021data} (blue solid square), rough surfaces from \cite{flack2020skin} (7 cases, green solid triangles), truncated cones from \cite{womack2022turbulent} (random: maroon triangles, staggered regular: maroon squares), 3 eggbox sinusoidal surface from \cite{rowin2024modelling} (yellow full circles), 31 cases from \cite{forooghi2017toward} (blue sideways open triangles),  3 power-law surfaces with exponent -1.5, -1 and -0.5 from \cite{barros2018measurements}  (green open triangles), surface with 7 cases of closely packed cubes (closed sideways red triangles) from \cite{xu2021flow}, and 4 cases of multiscale blocks  (open  sideways red triangles) from \cite{medjnoun2021turbulent}. Panel (a) shows the results in linear units, while panel (b) shows the ratio of modeled to measured sandgrain roughness in logarithmic units. }
   \vspace{-0.1in}
\label{fig:allsurftheta10}
\end{figure}

For each of the 104 surfaces, we first compute the wind-shade factor ${\cal W}_{\rm L}$ according to its definition in Eq. \ref{eq:defwindshade}. We use the turbulence spreading angle $\theta = 10$ degrees.  Then, with the known  experimental conditions for each surface expressed as the known $k_{\rm rms}^+$, we determine the effects of viscosity by finding $U_k^+$ according to  Eq. \ref{eq:ukplusiter}. The predicted roughness scale follows from Eq. \ref{eq:ksmod} using $a_p=3$.  The results are shown in Fig. \ref{fig:allsurftheta10} as scatter plots of predicted/modeled ($k_{\rm s-mod}$) versus measured ($k_{\rm s-data}$) sandgrain roughness scale, each normalized by $k_{\rm rms}$. The correlation coefficient is about 77\%. It is also useful to compare the average logarithmic error magnitude, i.e. $e=\langle | \log_{10}(k_{\rm s-mod}/k_{\rm s-data})|\rangle $.  The mean error is $e=0.16$ for the wind-shade model. 
The level of agreement between modeled and measured roughness scale over a large range of surface classes is encouraging.

\begin{figure}
\centering
    \begin{subfigure}{0.45\textwidth}
        \includegraphics[width=1.3\linewidth]{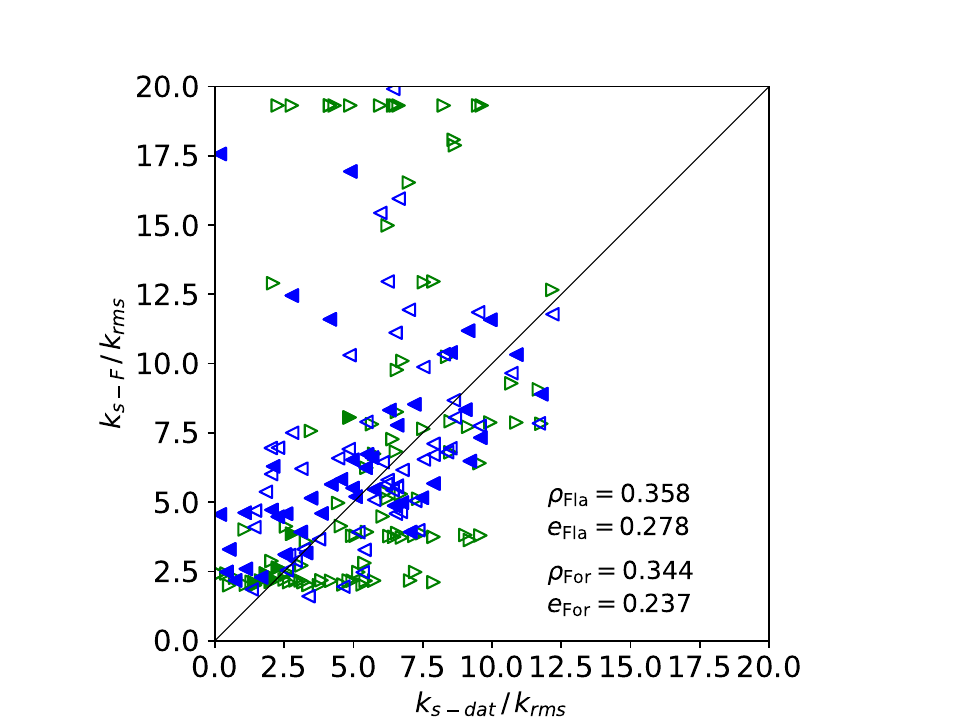}
        \caption{ }
    \end{subfigure} \hfill
     \begin{subfigure}{0.45\textwidth}
          \hspace{-0.4in}
        \includegraphics[width=1.3\linewidth]{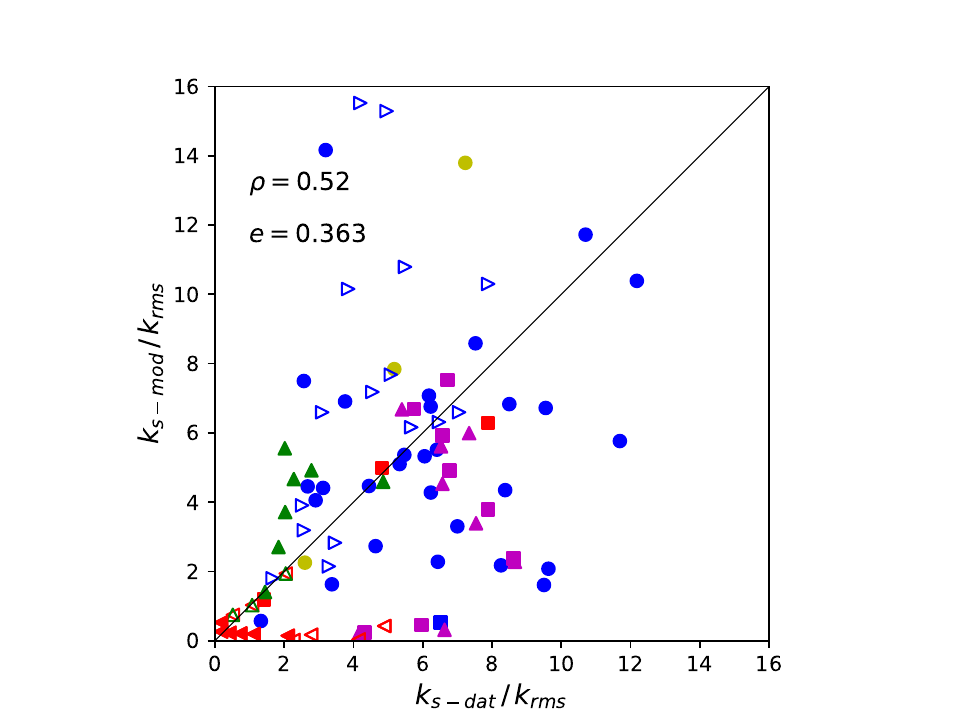}
        \caption{ }
    \end{subfigure}  
   \caption{(a) Sandgrain roughness predicted by the empirical  fits of  \cite{flack2020skin} 
  with truncated skewness (in green), and  \cite{forooghi2017toward} (in blue) versus measured values for all 104 surface cases considered in this work. Values used from the respective authors to fit the models are shown as solid symbols. (b) Sandgrain roughness predicted by the model without local pressure term, with $\theta=10^o$ and $a_p=7.5$ versus measured values for all 104 surfaces. Data and symbols same as in Fig. \ref{fig:allsurftheta10}.}
   \vspace{-0.1in}
\label{fig:allsurftheta10_Flacknopresshatx}
\end{figure}

For comparison with other common roughness models derived from particular datasets, in Fig. \ref{fig:allsurftheta10_Flacknopresshatx}(a) we show predictions using the correlations as listed in   \cite{flack2022important} and from \cite{forooghi2017toward}, referred to as $k_{\rm s-Fla}$ and $k_{\rm s-For}$, respectively. These models are implemented here as follows, 
\be
k_{\rm s-Fla} = 
k_{\rm rms} \, \left\{ \begin{array}{ll} 
  2.48  \, [1+({\rm min}(1.5,S_k))]^{2.24} & \quad    S_k >  0\\
  2.11        & \quad S_k = 0 \\
  2.73 \, [2+({\rm max}(-0.7,S_k))]^{-0.45} &  \quad  S_k > 0 
\end{array} \right.
\ee
\be
k_{\rm s-For} = k_t \,\,1.07\,[1-\exp(-3.5\, ES_x \,)]\,(0.67 S_k^2+0.93 S_k+1.3), 
\ee
where the first includes clipping of the skewness factor $S_k$ into the domain of validity of the fit, $ES_x=\langle |\partial h/\partial x| \rangle$ is the average slope magnitude, and $k_t$ is the samples' peak-to-trough scale, defined as $k_t={\rm max}(h)-{\rm min}(h)$. (This definition differs slightly from that used by 
\cite{forooghi2017toward} -- the latter included subdivision of the surface into parts whose extent is specified based on additional parameters -- but for present comparative purposes the effects on qualitative trends are relatively small). 
Over the 104 surfaces considered here, and including the arbitrary truncation mentioned above, the  correlation coefficients between measured and modeled sandgrain roughness lengths are 36\% and 34\%, respectively for these two models. The mean logarithmic errors are $e=0.28$   and $e=0.24$, respectively. We have verified that, as expected,  when restricting to data for which these correlations were originally developed and fitted by their authors, the correlation is significantly higher and the error lower  (see solid symbols in figure \ref{fig:allsurftheta10_Flacknopresshatx}a). 

Many other correlation and data-based models have been proposed (see review and summary in \cite{flack2022important}) but performing an exhaustive comparison with all of them  is beyond the scope of this paper. In addition, the various data-driven and machine learning approaches that have been proposed, while promising in principle, are more challenging to reproduce by others compared to simple function evaluations looped over the surface (as is required in the present approach to compute the wind-shade factor). The main contention of the present work is that our results show  that a single geometric parameter ${\cal W}_{\rm L}$ can be developed, and that inclusion of such a parameter (e.g.\ in any other correlations-based or ML based models) is beneficial, since  by itself, with only two adjustable parameters ($a_p$ and $\theta$), it already provides strong predictive power. 

\section{Analysis of model terms} \label{sec:analysis}

In this section we explore the effects of various included physical effects and model parameters. The effect of selecting $a_p=3$ is immediately obvious since it only serves as a multiplying factor leaving the correlation coefficient intact. However, selecting larger or lower $a_p$ (e.g. $a_p=3.5$ or $a_p=2.5$) leads to a cloud falling further above (or below) the line in Fig. \ref{fig:allsurftheta10} and slightly increases the error parameter $e$. An error fitting procedure shows that the minimum error is indeed obtained for $a_p=3.06$ but the error is almost unchanged. We conclude that selecting $a_p=3$ appears to be a good choice. 

Other model ingredients included the potential flow pressure distribution, i.e. selecting a local value of $\alpha(x,y)/(\pi+\alpha)$ inside the integral, using the projection of the square velocity via the term $\hat{n}_x^2(x,y)$, and including the shading factor $F^{\rm sh}(x,y)$.  The effect of viscous contributions can also be ascertained. Finally, a specified angle $\theta$ was chosen for the model. The effect of each of these choices is analyzed next and a summary of the resulting correlation and error coefficients are provided in tabular form (see Table \ref{table:corrs}).

\subsection{Potential flow pressure distribution}
To quantify the effect of the pressure distribution term, we here define a wind-shade factor using the overall mean pressure as prefactor instead of the local one and do not include the projection normal to the surface (i.e. we also omit the $\hat{n}_x^2$ term). We define

\be
{\cal W}^{\rm no \, p}_{\rm L} = \left<\frac{\alpha}{\pi+\alpha} \right> \, \left<    
\frac{\partial h}{\partial x}
 \, F^{\rm sh}(x,y;\theta)  \right> .
 \label{eq:windshadenopressure}
\ee 
Everything else is left unchanged and the sandgrain roughness is computed using Eq. \ref{eq:ksmod} but using ${\cal W}^{\rm no \, p}_{\rm L}$ instead of ${\cal W}_{\rm L}$. The `best' value of $a_p$ is obtained by minimizing the error and the resulting value $a_p=7.5$  is used. The scatter plot of predicted sandgrain roughness scales is shown in Fig. \ref{fig:allsurftheta10_Flacknopresshatx}(b). As can be seen, there is visibly much more scatter and the prediction has been severely degraded, with a significantly lower correlation coefficient of $\rho=0.52$ and larger error of $e=0.362$. We conclude that the physics introduced by considering the local pressure acting on surface segments at various angles $\alpha$ provides significant increased predictive power for roughness length estimation.  
 
\subsection{Effect of normal velocity projection}
We here examine the effect of the projection of the incoming velocity $U_k$ onto the direction normal to the surface given by the unit vector $\hat{\bf n}$ and its x-direction component. To isolate this factor, we compute the overall average of this term and use it to scale the average without it but assuming the mean of a random surface angle distribution for which $\langle n_x^2\rangle = 1/2$, i.e.

\be
{\cal W}^{\rm no \, nx}_{\rm L} =  \frac{1}{2} \, \left<  \frac{\alpha}{\pi+\alpha}   \, 
\frac{\partial h}{\partial x}
 \, F^{\rm sh}(x,y;\theta)  \right> .
 \label{eq:windshadenoproj}
\ee 
The sandgrain roughness is computed using Eq. \ref{eq:ksmod} but using ${\cal W}^{\rm no \,nx}_{\rm L}$ instead of ${\cal W}_{\rm L}$. To improve the agreement between this model and the data, the optimal parameter obtained by error minimization is $a_p=4.8$ (this change does not affect the correlation coefficient). The resulting scatter plot (not shown) is very similar to the baseline case and the correlation coefficient is still about $\rho=0.77$, and the error is similar to the baseline case, $e=0.168$.  For surfaces with differing strong anisotropic features the impact of velocity projection might be more noticeable. However, omission of the velocity projection does not seem to have any noticeably detrimental impact on the results, for the set of surfaces considered here. 

\subsection{Effect of near wall velocity profile}
\begin{figure}
\centering
    \begin{subfigure}{0.45\textwidth}
        \includegraphics[width=1.3\linewidth]{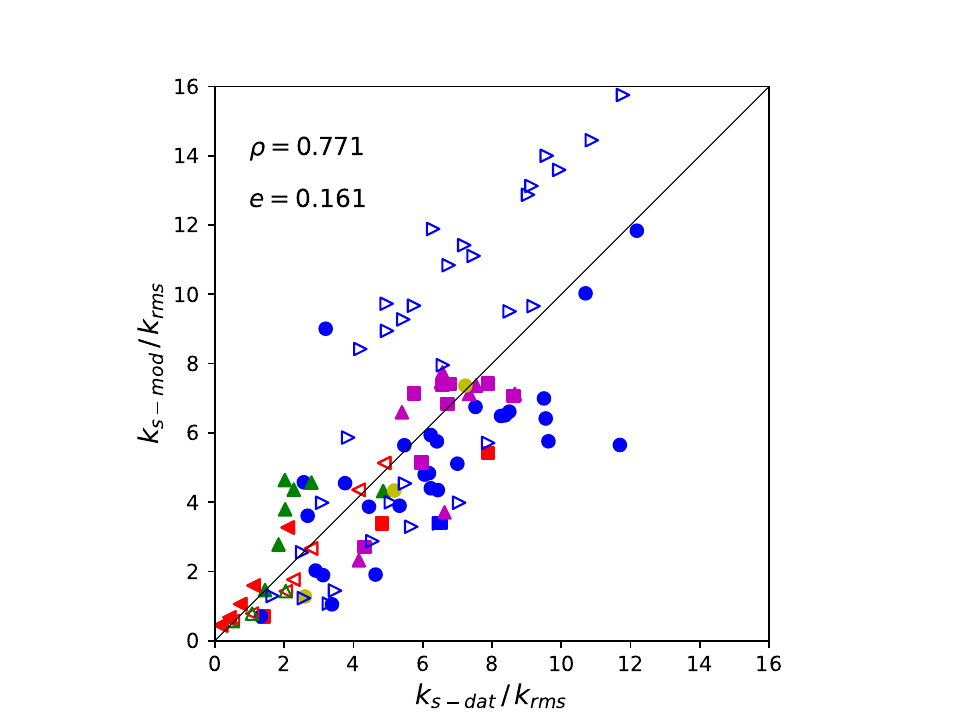}
        \caption{ }
    \end{subfigure} \hfill
     \begin{subfigure}{0.45\textwidth}
          \hspace{-0.4in}
        \includegraphics[width=1.3\linewidth]{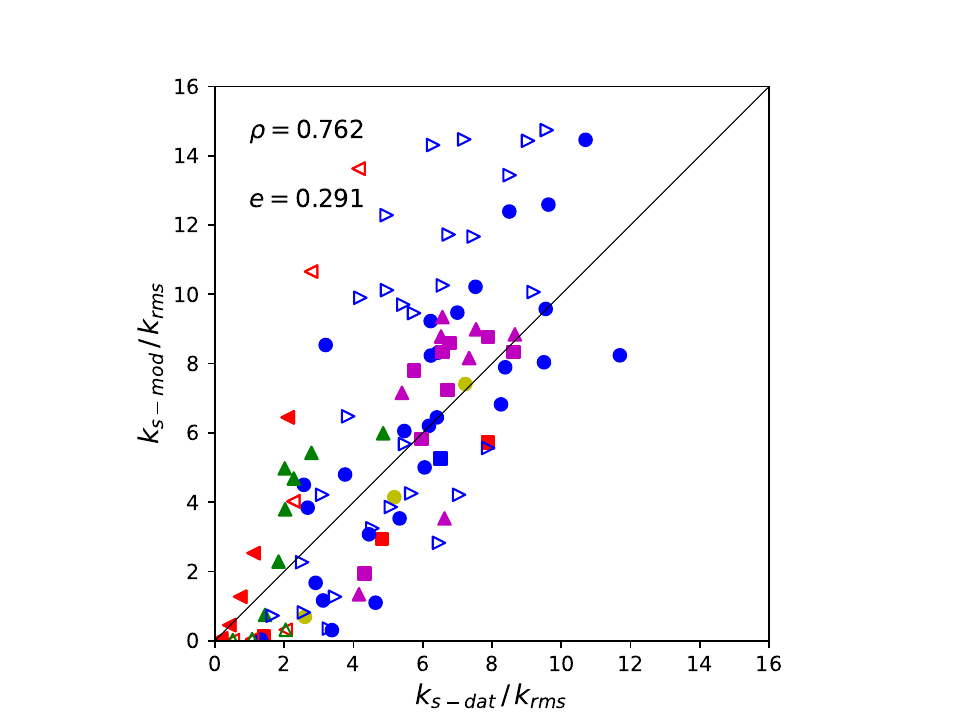}
        \caption{ }
    \end{subfigure}  
   \caption{(a) Sandgrain roughness predicted by the model including the velocity profile factor using the 1/7 power-law, for all 104 surfaces, and including the pressure and velocity projection case, using the optimal $a_p=4.5$. (b) Sandgrain roughness predicted by the baseline model but without the viscous term and setting using the optimal $a_p=3.8$. Data and symbols same as in Fig. \ref{fig:allsurftheta10}.}
   \vspace{-0.1in}
\label{fig:velprofnoviscous}
\end{figure}

We now test the effects of defining the wind-shade factor according to 
Eq. \ref{eq:defwindshadevel}, i.e. including a geometric factor that ranges between 0 and 1 between the lowest and highest surface points, with a 1/7 profile. A value of $a_p=4.5$ is used in order to provide a best fit to the data.  The results are shown in Fig. \ref{fig:velprofnoviscous}. The correlation coefficient and quality of the model is not improved and remains similar to the baseline case. For some of the datasets, noticeably the multiscale block surfaces \citep{medjnoun2021turbulent} and the closely spaced wall-attached cubes of \cite{xu2021flow}, the baseline model systematically overestimated $k_s/k_{\rm rms}$ while the results using an assumed 1/7 velocity profile are visibly better.  

However, overall the results are rather similar. We have experimented with other powers for the profile (1/2 and 2, and even the exponential  profile proposed by \cite{yang2016exponential}), but results became less good with such other choices. We conclude that disregarding the velocity profile factor is a good approach, at least when considering applicability to a wide range of surfaces. But, if specifically targeting surfaces with wall attached blocks (\cite{xu2021flow,medjnoun2021turbulent}, inclusion of the velocity profile factor, i.e. defining the wind-shade factor as ${\cal W}_{\rm L,vel}$ remains as a good option as well.

\subsection{Effects of viscosity for the datasets considered}
We have examined the effect of not including the viscous term in the prediction of $k_s$ (i.e. setting $c_f=0$). For this case $a_p=3.8$ is the optimal prefactor.  The results shown in Fig. \ref{fig:velprofnoviscous}(b) are visibly degraded especially for low roughness cases. While we obtain a similar correlation coefficient to the baseline case, of $\rho=0.76$, the model without the viscous drag results in a noticeably larger error of $e=0.286$. Again, when compared to the baseline model, results lead us to conclude that the inclusion of physics is beneficial to the model accuracy.  

\subsection{Turbulence spreading angle}\label{sec:TurbSpreadAngle}

To elucidate the sensitivity of the baseline case results to the assumed turbulence spreading angle, we apply the model with two other choices, namely $\theta = 5$ and 15 degrees. Scatter plots (not shown) have similar appearance to the baseline case, with correlation coefficients of 
0.689 and 0.753, respectively. Moreover, with the optimal values of
$a_p=9$ and $a_p=2.1$, respectively, we obtain errors of $e=0.19$  for  both the lower and larger angles. 

\begin{figure}
\centering
    \begin{subfigure}{0.45\textwidth}
        \includegraphics[width=1.3\linewidth]{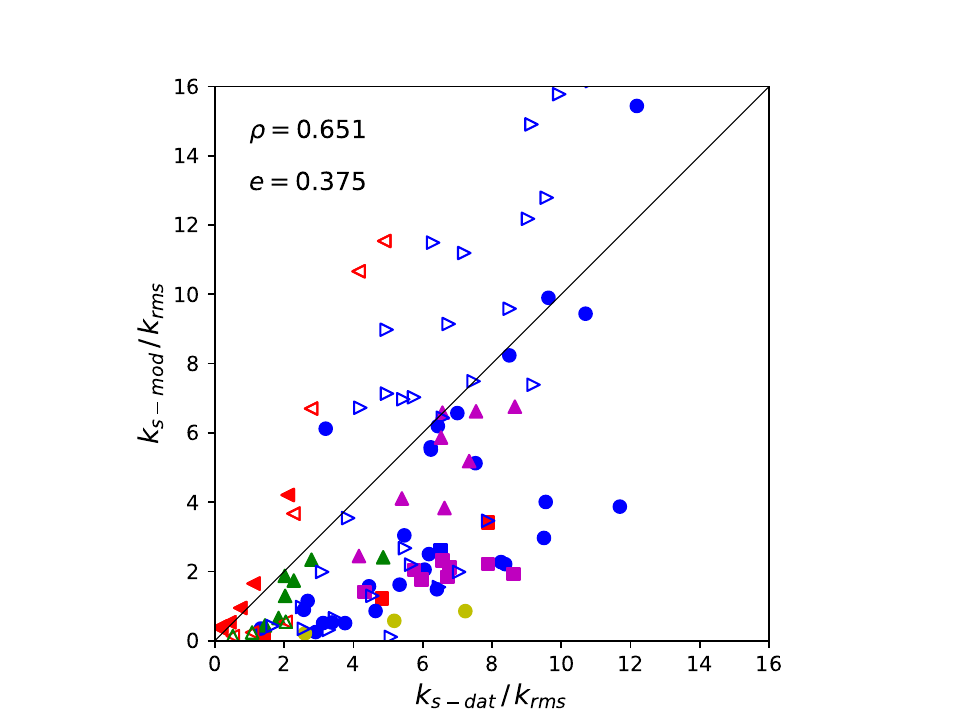}
        \caption{ }
    \end{subfigure} \hfill
     \begin{subfigure}{0.45\textwidth}
          \hspace{-0.4in}
        \includegraphics[width=1.3\linewidth]{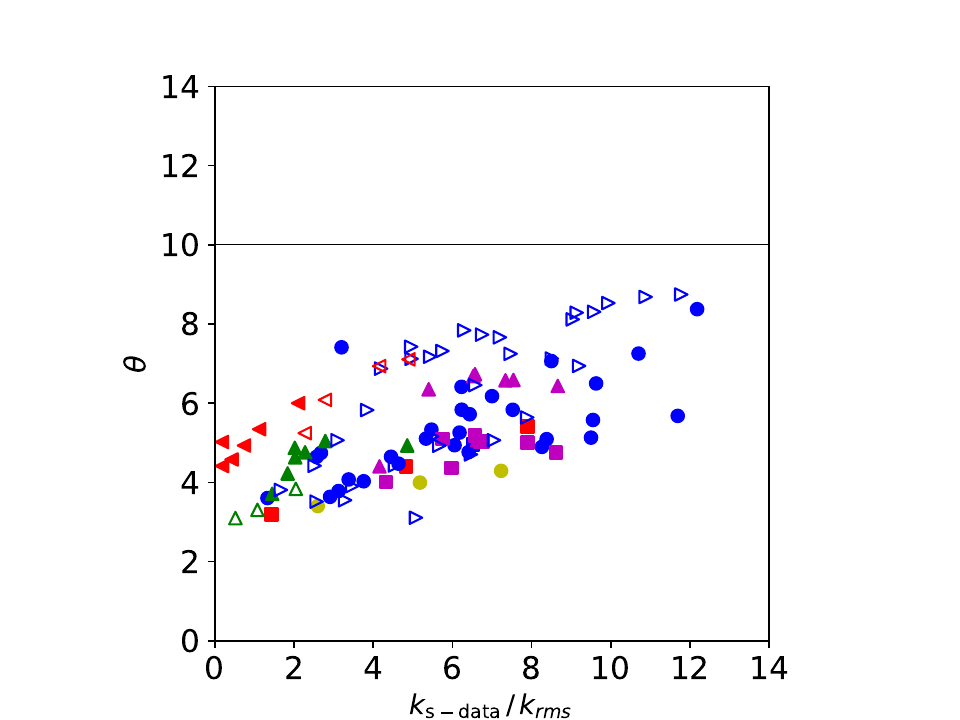}
        \caption{ }
    \end{subfigure}  
   \caption{(a) Sandgrain roughness predicted by the model with iterative determination of the turbulence spreading angle and $a_p=2.5$ versus measured values for all 104 surface cases considered in this work. (b) Turbulence spreading angle determined iteratively as part of the wind-shade model. Data and symbols same as in Fig. \ref{fig:allsurftheta10}.}
   \vspace{-0.1in}
\label{fig:allsurfdyntheta}
\end{figure}

Since there is some dependence on angle, we next test the idea of determining the angle $\theta$ from the model's ratio of turbulence to mean velocity and described by the ratio $u_\tau/U_k =\tan \theta$. The approach is iterative: we begin with an initial choice of $\theta=10^o$ and compute the shading function $F^{\rm sh}(x,y,\theta)$, the wind-shade factor and $U_k^+$ from Eq. \ref{eq:ukplusiter}, also including viscous effects.  Then $\theta=\arctan(1/U_k^+)$  is recomputed and the procedure iterated until the angles differ by less than $10^{-4}$. Remarkably the procedure converges rapidly, typically after less than 5-6 iterations. For this case, the prefactor $a_p=7.5$ was selected as the choice minimizing the error. The results are shown in Fig. \ref{fig:allsurfdyntheta}(a), while the iteratively determined angle is shown in Fig. \ref{fig:allsurfdyntheta}(b). As expected, the larger the roughness, the more turbulence intensity over mean velocity at $z_p$ and thus a larger spreading angle is predicted, and vice-versa. However, the spread has increased and the correlation and the error are now $\rho=0.651$ and $e=0.34$ compared to the baseline model, respectively and also worse than for the two fixed other angles tested. This result leads us to the conclusion that fixing a single angle is likely to be a better approach and solving dynamically for the angle has not increased the model's predictive power. 

Since from Fig. \ref{fig:allsurfdyntheta}(b) it appears that the average angle is closer to 6 degrees than 10 degrees, the calculation over all surfaces is repeated for 6 degrees. The results (correlation and error) are almost the same as those listed above for 5 degrees (and we note that the value of $a_p \sim 9$ is rather large (it would reach significantly above the roughness sublayer) and would appear less physically justified than the chosen baseline value of $a_p=3$). As can be seen, the sensitivity of the model prediction quality to the turbulence spreading angle is rather weak and 10 degrees appears a good a-priori choice and is therefore kept as the baseline model since it led to the largest correlation coefficient and smallest mean logarithmic error. 

\subsection{Some other options}

In this work we have introduced a new lengthscale $k_p^\prime$ corresponding to a measure of positive-only deviations from the mean height. As such, the question arises whethere it could, by itself, serve as a good model parameter by setting, e.g. the sandgrain roughness proportional to this scale according to
\be
k_s = a_p \, k_p^\prime.
\label{eq:kpprimeonly}
\ee
In this case (scatter plot not shown) there is very little correlation with the roughness length (it is found  to be $\rho = 0.312$). Clearly, this simple definition of a height is by itself insufficient as a model. 

Finally, we examine the effect of surface slope by itself. We do not take into account the shading factor and simply take the average of slope where it is positive (i.e. only accounting for forward facing portions of the surface but without including the shading nor pressure distribution and velocity projection). This amounts to computing the factor as
\be
{\cal W}^{\rm no \, shade}_{\rm L} =  0.025 \,\, \left<    
 R\left(\frac{\partial h}{\partial x}\right) \right> .
 \label{eq:noshade}
\ee 
where $R(x)$ is the ramp function. The  prefactors for ${\cal W}^{\rm no \, shade}_{\rm L}$ of 0.025, and $a_p=4$, are selected to minimize the average error in this case.  The correlation is only $\rho=0.384$ and the error is $e=0.33$, so clearly also this approach does not represent the correct trends. While the  factor ${\cal W}^{\rm no \, shade}_{\rm L}$ does not contain shading effects, it is important to recognize that this model still contains some qualitative shading features through its proportionality to the scale $k_p^\prime$: when the surface is very negatively skewed (e.g. pitted with valleys typically shading the downstream portion), $k_p^\prime$ is relatively small as opposed to when the surface is positively skewed. Still,  the model is insufficient to reproduce realistic trends. 

Finally, we have examined dependence of model predictions on the choice of $p=8$ for the high-order moment used to identify the peak positive heigh deviation. Model evaluations using $p=12$ and $p=6$ yielded the same results as using $p=8$, specifically, correlation coefficients and mean logarithmic errors differences of less than 0.5\%.  

Table \ref{table:corrs} summarizes the correlation coefficients and mean logarithmic errors for each of the cases tested. It is apparent that the baseline model with uniform or 1/7 velocity profile term display the largest correlation coefficient and lowest mean logarithmic error. 

\begin{table}
\vskip -0.2in
\centering
\caption{Summary of correlation coefficients and mean logarithmic error between model predicted and measured sandgrain roughness for various versions of the model}
\begin{tabular}{|c|c|c|}
\hline
Model version & corr. $\rho(k_{\rm s.mod},k_{\rm s.dat})$ &  
error: $e= \langle \log_{10} |\frac{k_{\rm s.mod} }{ k_{\rm s.dat}}|\rangle$ \\ \hline
Baseline model, Eqs. \ref{eq:ksmod},\ref{eq:defwindshade},\ref{eq:ukplusiter}, $\theta=10^o$:
& 0.77 & 0.16 \\ \hline
Without pressure, Eqs. \ref{eq:ksmod},\ref{eq:windshadenopressure},\ref{eq:ukplusiter}
& 0.52 & 0.36\\ 
\hline
No normal projection,  Eqs. \ref{eq:ksmod},\ref{eq:windshadenoproj},\ref{eq:ukplusiter}  & 0.77& 0.17\\
\hline 
 With velocity profile, Eqs. \ref{eq:ksmod},\ref{eq:defwindshadevel},\ref{eq:ukplusiter}  & 0.77 & 0.16 \\
\hline 
 Without viscous effects, Eqs. \ref{eq:ksmod},\ref{eq:defwindshade},\ref{eq:ukvswl} & 0.76 & 0.29 \\
\hline 
 Baseline model, with $\theta=5^o$  & 0.69 & 0.19 \\
\hline 
 Baseline model, with $\theta=15^o$  & 0.75  & 0.19 \\
\hline
Baseline model, $\theta=\arctan(1/U_k^+)$ & 0.65  & 0.34 \\
\hline
Simple max positive height, Eq. \ref{eq:kpprimeonly}   & 0.21  & 1.41 \\
\hline
Positive slope only, no shading, Eq. \ref{eq:noshade}   & 0.38  &  0.33\\
\hline

\end{tabular}
\label{table:corrs}
\end{table}

\section{Results for transitionally rough conditions}
In order to illustrate the predictive power of the wind-shade model also in the transitionally rough regime, we apply it for any given surface over a large range of values of $k_{\rm rms}^+$, from $k_{\rm rms}^+\to 0$ (fully smooth) to very large values $k_{\rm rms}^+ \to \infty$ (fully rough). Results are shown in the form of the familiar roughness function $\Delta U^+(k^+_{s\infty})$. The roughness function (velocity deficit) is expressed in terms of the asymptotic (fully rough) equivalent sandgrain roughness 
$k^+_{s\infty}$. For a given $k^+_{s\infty}$, we determine the corresponding nominal model height in wall units, $z_k^+$, by applying the model in the fully rough limit and obtain
\be z_k^+ = k_{s\infty}^+ \, \exp[\kappa( {\cal W}_{\rm L}^{-1/2}- 8.5)].
\label{eq:zkks}
\ee

For a given $z_k^+$, Eq. \ref{eq:mombalance} has to be solved for $U_k^+$. Once we know  $U_k^+=U^+(z_k^+)$, we may determine the roughness function $\Delta U^+$ from its difference with the velocity profile for a smooth wall.  To ensure that $\Delta U^+$ tends to zero for small $k_{s\infty}^+$, when $z_k^+$ becomes small (e.g. approaching the viscous sub-layer), the smooth wall velocity profile cannot be assumed to be logarithmic and needs to merge with linear near-wall behavior.  We use the velocity profile transition function proposed in \cite{fowler2022lagrangian} (Eq. C1), so that the model prediction for the roughness function becomes:
\be
 \Delta U^+(k^+_{s\infty}) = \left(\frac{1}{\kappa} \log (\kappa_2+z_k^+) + B  \right)\,\left(1+(\kappa_1^{-1} z_k^+)^{-\beta}\right)^{-1/\beta}\,-\,U_k^+,
\ee
where $z_k^+$ is given in terms of the prescribed $k^+_{s\infty}$ by Eq. \ref{eq:zkks}. The first factor in the fit \citep{fowler2022lagrangian} is the smooth-wall profile in the logarithmic region, while the second factor ensures a linear behavior near zero. Fitting parameters are given by $\kappa=0.4$, $B=5$, $\kappa_2=9.753$, $\beta=1.903$, $\kappa_1=\kappa^{-1} \log \kappa_2 + B$.  
\begin{figure}
\centering
    \begin{subfigure}{0.45\textwidth}
        \includegraphics[width=1.2\linewidth]{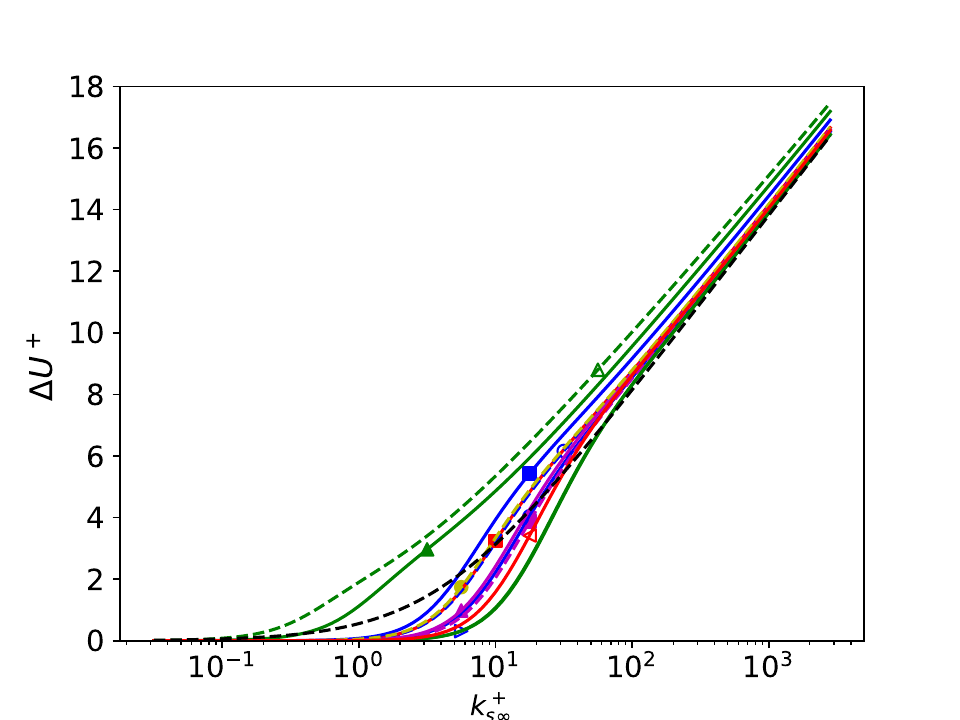}
        \caption{ }
    \end{subfigure} \hfill
     \begin{subfigure}{0.45\textwidth}
          \hspace{-0.3in}
        \includegraphics[width=1.2\linewidth]{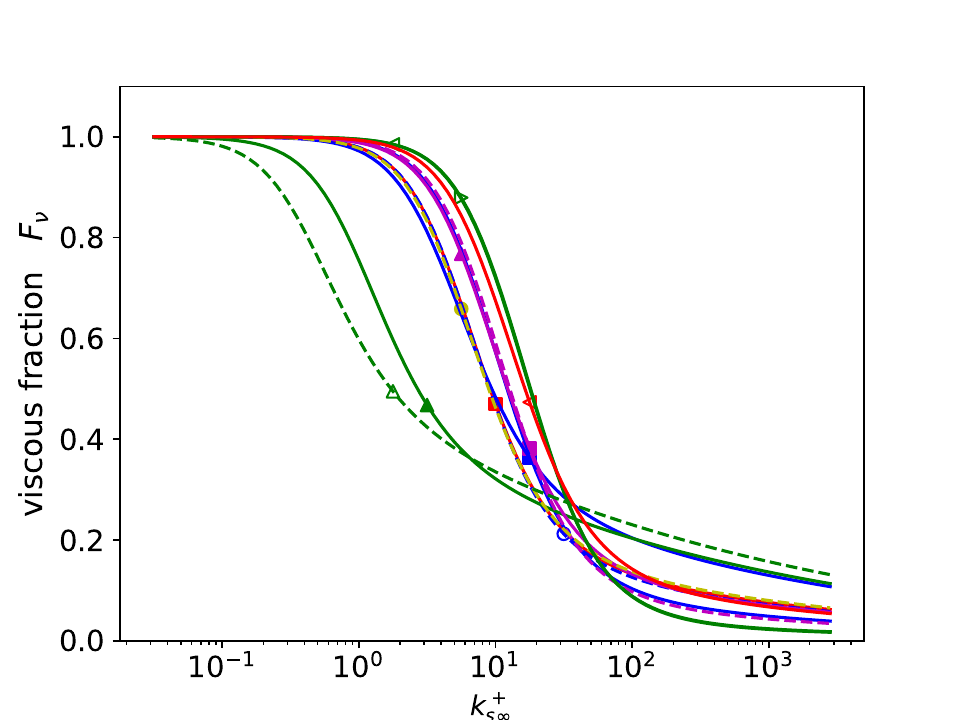}
        \caption{ }
    \end{subfigure}  
   \caption{(a) Velocity deficit roughness function predicted by the windshade roughness model for the 12 sample surfaces shown in Fig. \ref{fig:allfigs} as function of reference (fully rough) sandgrain roughness $k_{s\infty}^+$. (b) Fraction of viscous over total drag as function of reference (fully rough) sandgrain roughness $k_{s\infty}^+$ as predicted by the wind-shade roughness model for the 12 sample surfaces. The black dashed line is the Colebrook formula: $\Delta U^+_{\rm Cb} = 2.5 \, \ln(1+0.25 k_{s\infty}^+)$. The 12 surfaces are those of Fig. \ref{fig:allfigs}: Surface are from: \cite{jelly2022impact} (red solid line and full red squares), \cite{jouybari2021data} (sandgrain-type: blue line and solid circles, sinusoidal: dashed blue line and open circle, cubes: solid blue line and solid blue square), rough surface (case 1) from \cite{flack2020skin} (green line and solid triangles, 
   truncated cones from \cite{womack2022turbulent} (random: maroon line and triangles, staggered regular: maroon line and squares), eggbox sinusoidal surface from \cite{chung2021predicting} (yellow dashed line and full circle), two surfaces from \cite{forooghi2017toward} (blue sideways triangles),  power-law surface  with exponent -0.5 from \cite{barros2018measurements}  (green dashed line and open triangle), and the multiscale (iter123) Lego block surface from \cite{medjnoun2021turbulent} (red line and sideways red triangle).}
   \vspace{-0.1in}
\label{fig:deltaUfraction}
\end{figure}

\begin{figure}
\centering   
    \begin{subfigure}{0.45\textwidth}
        \includegraphics[width=1.2\linewidth]{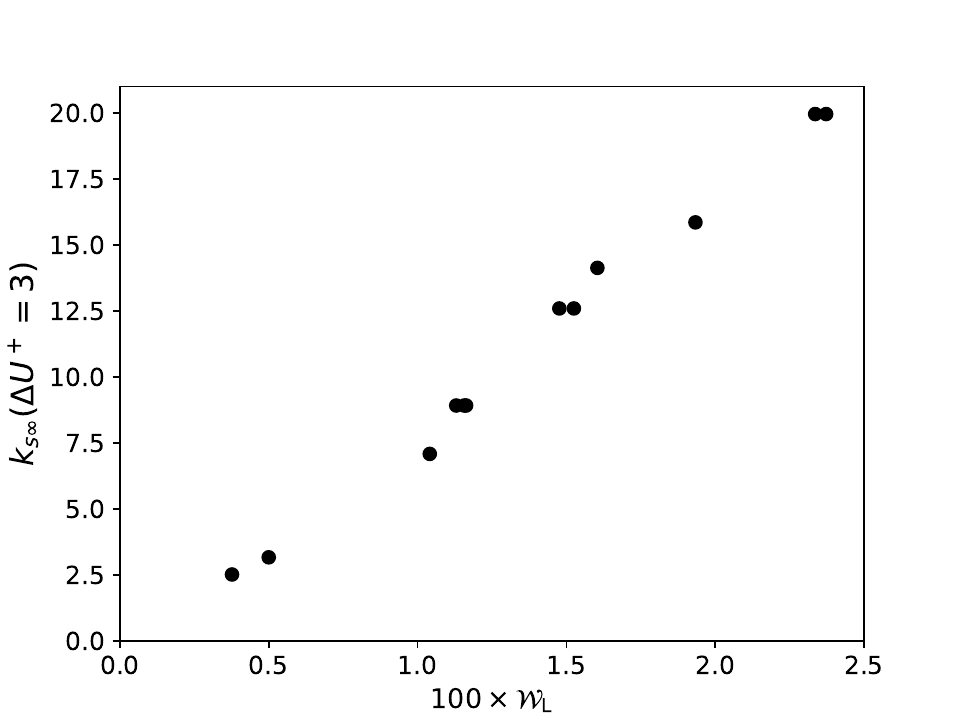}
        \caption{ }
    \end{subfigure} \hfill
     \begin{subfigure}{0.45\textwidth}
          \hspace{-0.3in}
        \includegraphics[width=1.2\linewidth]{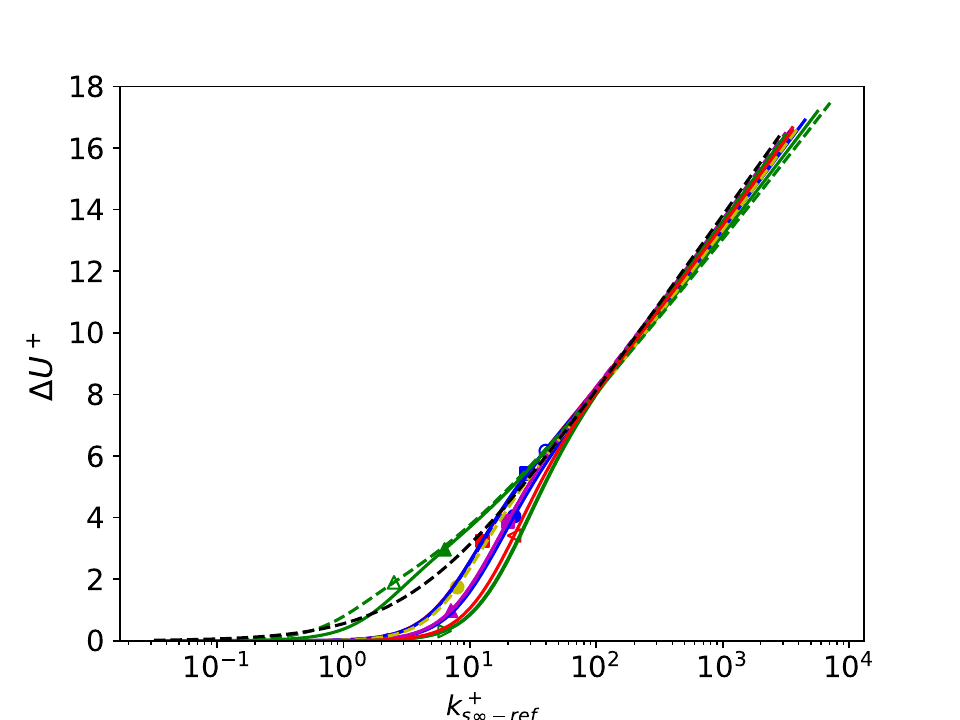}
        \caption{ }
    \end{subfigure} 
   \caption{(a) Effective asymptotic roughness scale $k_{s\infty}^+$ at which $\Delta U^+=3$ as function of wind shade factor, showing strong correlation. (b) Same as Fig. \ref{fig:deltaUfraction} but with $k_{s\infty{\rm -ref}}^{+}$ determined by forcing the lines to the fully rough regime at $\Delta U^+=8.14$ (or up to $k_{s\infty}^+=100$ according to the Colebrook formula). Lines and symbols same as in Fig. \ref{fig:deltaUfraction}.}
   \vspace{-0.1in}
\label{fig:deltaUvsksinfref}
\end{figure}

As a representative sample, in figure \ref{fig:deltaUfraction} we apply this method to the 12 surfaces shown in Fig. \ref{fig:allsurftheta10}.   Figure \ref{fig:deltaUfraction}(a), which shows the model predicted roughness function $\Delta U^+$ against $k_{s\infty}^+$ demonstrates that the model is capable of capturing Colebrook and Nikuradse type behaviours in the transitional regime. It remains to be determined whether these trends are fully consistent with the dataset, but we note that the model correctly captures the Colebrook like tendency of the `Data1' surface of \cite{flack2020skin} and the power-law surface  with exponent -0.5 from \cite{barros2018measurements}. Also shown (in Fig. \ref{fig:deltaUfraction}(b)) is the model predicted friction drag fraction for the 12 representative surfaces. This fraction is evaluated according to

\be
{\cal F}_\nu(Re_k) = \frac{\frac{1}{2} c_f(Re_k) \,\overline{F}_\nu^{\rm sh}}{{\cal W}_L+\frac{1}{2}c_f(Re_k)\,\overline{F}_\nu^{\rm sh}},
\ee
which is plotted as function of $k_{s\infty}^+$.  The impact of friction is noteworthy, even at elevated $k_{s\infty}^+$ extending well above 100. This observation is consistent with direct measurements of drag ratios as discussed \cite[][figure 10]{busse2017reynolds},  \cite[][ figure 10]{macdonald2018direct}, \cite[][figure 7\textit{d}]{jelly2022impact}\cite[][figure 4]{chan2014numerical}, all showing that at $\varDelta U^+ \approx 6$--$7 \Leftrightarrow k_{s\infty}^+\approx 50$--$70$, the viscous--pressure drag partition is still as high as $30$--$40\%$.  A comparison of figure \ref{fig:deltaUfraction}(a) and (b) indicates that those surfaces exhibiting Colebrook type transitional behaviour, are also those with a slower decay in the viscous drag partition with increasing $k_{s\infty}$. 
There exists strong correlation also with the windshade factor.
In  Figure \ref{fig:deltaUvsksinfref}(a) we show the asymptotic roughness length $k_{s\infty}$ at which $\Delta U^+=3$, an indication of Colebrook type (low  $k_{s\infty}(\Delta U^+=1)$) or more Nikuradse-type (higher $k_{s\infty}(\Delta U^+=1)$) behavior, as function of the windshade factor. A  strong correlation is apparent. 
In general, the behaviours shown in figure\nhu{s \ref{fig:deltaUfraction} and \ref{fig:deltaUvsksinfref}} suggest that the wind-shade model could provide a useful sandpit for investigating the transitional regime.

For reference, we can also mimic the customary approach of assuming that at some large value of $k_{s\infty}^+$ the behavior should be that of hydrodynamically fully rough asymptotics. We thus shift all the curves and redefine a new $k_{s\infty-{\rm ref}}^+$ such that the curves agree with the Colebrook formula at 
$k_{s\infty}^+=100$ (or at $\Delta U^+=8.14$). 
Figure \ref{fig:deltaUvsksinfref}(b) displays the result. The curves suggest that, strictly speaking, many of the surfaces have not truly reached their fully rough asymptotes at $k_{s\infty-{\rm ref}}^+= 100$, consistent with the slow decay of the viscous-drag fraction in figure~\ref{fig:deltaUfraction}(\textit{b}).
The errors extrapolated to $k_{s\infty-{\rm ref}}^+= 10^3$--$10^4$, however, remain limited to about one unit of $\varDelta U^+$ only.

\section{Conclusions}
\label{sec:conclusions}

A model is proposed that can predict the drag over arbitrary rough surface topographies, hence permitting the prediction of roughness length ($k_s$ or $z_0$) directly from surface topography. The model is built around a flow-physics motivated parameter called the wind-shade factor which accounts for the effect of sheltering. The intention is to consolidate the physics behind topographical parameters such as the frontal and plan solidities, and effective slope and skewness, as well as to address the effects of roughness clustering and directionality. Sheltered and unsheltered regions are identified using a backward horizon function and an assumed flow separation angle $\theta$. The pressure force acting on the exposed (unsheltered) regions of the surface topography is calculated using a piece-wise potential flow approximation for a ramp with the local surface gradient \citep[as previously used to model the surface drag of free-surface waves by][] {ayala2024moving}. Recognising that even in the apparent fully-rough limit a substantial proportion of the overall wall drag remains attributed to viscous wall shear stress, we also include a viscous drag model which approximates the wall shear stress in the unsheltered region using a simple friction factor approach. The model is tuned and tested against a database of 104 surfaces, most of them being freely available from a public database 
(\href{https://roughnessdatabase.org}{\color{blue}https://roughnessdatabase.org}).
To avoid over-fitting, simple flow physics-based modeling assumptions are preferred. However, certain modeling parameters remain; namely, the turbulence spreading (or flow separation) angle $\theta$, the edge of the roughness sublayer $a_p$ and in one model variant, the form of the mean velocity profile within the roughness sublayer $U(z)$. On these three parameters, we are guided by the literature. The accuracy of the model predictions is assessed by comparing the model predicted equivalent sandgrain roughness $k_{\rm s-mod}$ against that reported in the database $k_{\rm s-data}$ for the {104} surfaces. The accuracy is quantified using the the overall correlation $\rho$ (which ideally would be close to 1) and the average error magnitude $e$ (which should be minimised). In it's baseline form (using estimates for $\theta$ and $a_p$ from literature), the model gives $\rho = 0.77$ and $e = 0.16$. This is a substantial improvement over topographical curve fit models of \cite{flack2022important} and \cite{forooghi2017toward}, which both yield substantially lower $\rho$ and higher $e$. In fairness, this might be expected since these models were originally proposed on a more limited database that does not cover the wide parameter space of the many surface classes considered here. The important point though is that with minimal tuning, based on readily assumed constants from the literature, the physics-based approach of the wind-shade model is capable of capturing the salient behaviour across the selected database, outperforming existing topographical fits. 

The functionality of the model is investigated by varying the modeling variables ($\theta$, $a_p$ and $U(z)$) and also deactivating different physics components in the model. In its baseline form, $a_p = 3$ is found to give the optimal prediction (in line with standard estimates for the edge of the roughness sublayer). Changing the turbulence spreading angle from the optimal $\theta = 10^\circ$ to $5^\circ$ or $15^\circ$ causes marginal degradation in the quality of the prediction (reduced $\rho$, increased $e$). A refinement is also tested where the turbulence spreading angle is dynamically set based on the local wall normal fluctuation magnitude (see \S~\ref{sec:TurbSpreadAngle}). This refinement in $\theta$ diminishes the quality of the predictions, and it is found that best results are obtained taking a fixed angle of $\theta \approx 10^\circ$. Changing the near wall velocity profile from assumed plug flow to a $1/7$ power law makes almost no difference in predictive accuracy, and for now the plug flow model is preferred for simplicity (but with the recognition that for certain multiscale surfaces the power-law profile may have advantages).

The principal assumption of the wind-shade model is that the drag can be determined from the pressure difference between the front exposed face of the roughness and the back sheltered region. This is computed by projecting the oncoming flow onto the surface normal of the exposed surface, and using a potential flow model to compute the mean pressure on each ramp-like element. The effect of this surface normal projection is tested by negating this projection, and instead computing pressure based on the streamwise velocity $U_k$ scaled with a prefactor based on the average surface normal projection for a random rough surface ($\langle n_x^2 \rangle = 1/2$). It is found that this step has no detrimental impact on the modelling accuracy. Conversely, when the model is altered from considering a local pressure (based on local slope) to a mean pressure computed over the entire surface (and a mean shade factor), this over-simplification leads to a substantial degradation in the predictive power of the model, indicating that consideration of the local pressure force acting on the surface segments, at various angles, is a critical component of the model. This failure of the mean pressure prediction further shows that a purely topographical approach to sheltering, based on computing the mean effective slope only from the exposed regions has more limited predictive capability.

Finally, the effect of the viscous drag component of the model is isolated by deactivating this term and comparing predictions. Without the viscous term ($c_f = 0$), the predictive correlation $\rho$ remains similar, but with an increase in mean error $e$. It is concluded that the inclusion of the viscous drag physics in the exposed (unshaded) regions is beneficial to the model accuracy. Moreover, it is demonstrated that the inclusion of the viscous term permits meaningful investigation of the transitionally rough regime. The viscous model is able to capture both Colebrook and Nikuradse type behaviours in the transitional trend of roughness function $\Delta U^+$ as a function of $k_s^+$, suggesting that this model might be able provide insight into the important and poorly understood transitional regime. From limited investigations, we can see that the viscous term is able to reproduce the Colebrook and Nikuradse type behaviours reported for certain roughnesses in the database, such as the Colebrook behavior seen in the data (set Data 1) of \cite{flack2020skin}.

The present modeling approach opens many avenues for further research and analysis. The model should be tested on additional surfaces and roughness classes. It could be combined with data-driven approaches to further improve the model's predictive abilities, making sure over-fitting is avoided and that resulting prediction tools can be broadly used by others. And, further extensions could be examined, such as modeling transverse forces from highly anisotropic surfaces. 

\section*{Acknowledgements}
The authors gratefully acknowledge the roughness database hosted at the University of Southampton that provides public access to crucially important data in this field. They also thank Dr. T. Jelly for help accessing the surface data for cases \#1,2,3, and Prof. I. Marusic for valuable comments and encouraging remarks about this project. CM acknowledges the Miegunyah fellowship  and the Department of Mechanical Engineering at the University of Melbourne for making a sabbatical visit there possible, as well as partial support from the Office of Naval Research (grant \# N00014-21-1-2162) and AFOSR (grant \# FA9550-23-1-0269)). DC and NH acknowledge the support of the Air Force Office of Scientific Research under award number FA2386-23-1-4071 (program manager: D.J. Newell, AOARD)  and also support from the Australian Research Council via the Discovery and Linkage Programs.

\section*{Declaration of interests} The authors report no conflict of interest.

\appendix

\vfill
\newpage

\section*{Appendix A: Datasets and roughness database}

In this appendix we provide listing of all the datasets analyzed in this paper. Tables \ref{table:full_data1} and \ref{table:full_data2} lists surface names and the surfaces $k_{\rm rms}$ value in viscous units for the experimental conditions considered. Also listed are values of $k_s/k_{\rm rms}$ from the data and from the wind-shade roughness model. The last column list the computed wind-shade factor ${\cal W}_{\rm L}$. 

The surfaces considered include 3 surfaces from \cite{jelly2022impact} (isotropic, ridges aligned in the streamwise ($x$), and aligned in transverse ($y$) directions), and 30 surfaces from the extensive study by \cite{jouybari2021data} catalogued in that study as fully rough. They were cases C07-C24 and C31 are for surfaces built from random ellipsoids using the method of \cite{scotti2006direct}; case C40 is for random Fourier mode surface, cases C26-C30 for sinusoidal waves of various wavelengths of same amplitude and differing slopes, C04-C06 for regularly arranged ellipsoids, C14-C18 for regularly placed ellipsoids in a vertical orientation, C43: random sandgrain roughness, C44: turbine blade rough surface, C45: wall attached cubes. More details can be found in \cite{jouybari2021data}. The there are 7 surfaces from \cite{flack2020skin} that were generated via random-phase Fourier mode superposition to create various slope and skewness parameters. Water channel measurements were used for flow characterization. Using the same experimental facility there are 16 surfaces consisting of truncated cones \citep{womack2022turbulent}. 8 of these (surfaces R10-R78) were in random arrangements with increasing density, while 8 were in regular, staggered arrangement (surfaces S10-S78). Three two-dimensional sinuosoidal surfaces with different steepnesses numerically studied by \cite{rowin2024modelling} are included. 60 surfaces studied numerically by \cite{forooghi2017toward} include various shapes leading to a range of skewnesses and element shapes. Three surfaces \cite{barros2018measurements} were generated using power-law spectra and random phases, and included smoother (power-law exponent $p=-1.5)$, intermediate $p=-1$, and rough $p=-0.5$ cases. 6 surfaces with wall attached, closely spaced, cubes studied numerically in \cite{xu2021flow} are also considered. Finally, 4 surfaces of blocks of decreasing sizes with varying number of iterations studied experimentally \cite{medjnoun2021turbulent} are included. 

The elevation maps for these surfaces are obtained from the roughness data-base 
(\href{https://roughnessdatabase.org}{\color{blue}https://roughnessdatabase.org}), except for the 6 surfaces from \cite{jelly2022impact} and \cite{rowin2024modelling} that were available from in-house sources.

\begin{table}
\vskip -0.2in
\centering
\caption{Surfaces, properties and wind-shade roughness model predictions. }
\begin{tabular}{|c|c|c|c|c|c|c|}
\hline
Case \# & Dataset name & $k_{\rm rms}^+$ & $k_{\rm s-data}/k_{\rm rms}$ & $k_{\rm s-mod}/k_{\rm rms}$ & ${\cal W}_{\rm L} \times 100$ \\ \hline
0 & Jelly-surf-iso & 27.0 & 4.83 & 2.941 & 0.854 \\
1 & Jelly-surf-aniso-x & 27.0 & 1.425 & 0.643 & 0.281 \\
2 & Jelly-surf-aniso-y & 27.0 & 7.89 & 4.912 & 1.157 \\
3 & Yuan-2021-C07 & 22.0 & 6.18 & 5.031 & 1.344 \\
4 & Yuan-2021-C08 & 30.0 & 10.7 & 10.201 & 2.142 \\
5 & Yuan-2021-C09 & 21.0 & 6.23 & 7.0 & 1.696 \\
6 & Yuan-2021-C10 & 30.0 & 11.69 & 6.34 & 1.487 \\
7 & Yuan-2021-C11 & 44.0 & 12.18 & 12.598 & 2.411 \\
8 & Yuan-2021-C12 & 32.0 & 8.5 & 8.909 & 1.989 \\
9 & Yuan-2021-C19 & 21.0 & 7.524 & 7.871 & 1.525 \\
10 & Yuan-2021-C20 & 17.0 & 6.235 & 7.055 & 1.286 \\
11 & Yuan-2021-C21 & 16.0 & 6.437 & 7.634 & 1.161 \\
12 & Yuan-2021-C22 & 29.0 & 9.55 & 7.25 & 1.525 \\
13 & Yuan-2021-C23 & 25.0 & 7.0 & 7.65 & 1.43 \\
14 & Yuan-2021-C24 & 27.0 & 9.629 & 10.076 & 1.481 \\
15 & Yuan-2021-C31 & 11.0 & 4.45 & 3.483 & 0.881 \\
16 & Yuan-2021-C37 & 18.0 & 6.055 & 4.416 & 1.138 \\
17 & Yuan-2021-C40 & 16.0 & 3.125 & 1.694 & 0.583 \\
18 & Yuan-2021-C26 & 14.0 & 4.64 & 1.769 & 0.672 \\
19 & Yuan-2021-C28 & 14.0 & 3.38 & 0.994 & 0.42 \\
20 & Yuan-2021-C29 & 21.0 & 5.333 & 3.258 & 1.163 \\
21 & Yuan-2021-C30 & 21.0 & 3.2 & 6.458 & 2.101 \\
22 & Yuan-2021-C04 & 22.0 & 2.91 & 1.859 & 0.822 \\
23 & Yuan-2021-C05 & 33.0 & 3.76 & 3.794 & 1.346 \\
24 & Yuan-2021-C06 & 22.0 & 2.68 & 3.358 & 1.219 \\
25 & Yuan-2021-C14 & 22.0 & 6.41 & 5.248 & 1.432 \\
26 & Yuan-2021-C15 & 19.0 & 8.26 & 6.105 & 1.248 \\
27 & Yuan-2021-C16 & 30.0 & 2.57 & 3.611 & 1.453 \\
28 & Yuan-2021-C17 & 31.0 & 8.38 & 6.193 & 1.535 \\
29 & Yuan-2021-C18 & 26.0 & 9.5 & 6.843 & 1.328 \\
30 & Yuan-2021-C43 & 17.0 & 5.47 & 5.201 & 1.228 \\
31 & Yuan-2021-C44 & 18.0 & 1.33 & 0.581 & 0.214 \\
32 & Yuan-2021-C45 & 23.0 & 6.52 & 5.491 & 1.041 \\
33 & Flack-2020-Data1-Tile1 & 22.7 & 1.45 & 1.306 & 0.5 \\
34 & Flack-2020-Data2-Tile1 & 43.2 & 2.28 & 3.857 & 1.094 \\
35 & Flack-2020-Data3-Tile1 & 41.2 & 1.84 & 2.158 & 0.916 \\
36 & Flack-2020-Data4-Tile1 & 41.6 & 2.03 & 3.188 & 1.068 \\
37 & Flack-2020-Data5-Tile1 & 42.9 & 2.79 & 4.323 & 1.167 \\
38 & Flack-2020-Data6-Tile1 & 49.3 & 4.86 & 4.702 & 1.147 \\
39 & Flack-2020-Data7-Tile1 & 47.8 & 2.02 & 3.993 & 1.152 \\
40 & Womack-2022-R10 & 34.8 & 4.16 & 2.529 & 0.475 \\
41 & Womack-2022-R17 & 49.3 & 6.63 & 3.938 & 0.756 \\
42 & Womack-2022-R39 & 75.5 & 8.66 & 6.785 & 1.476 \\
43 & Womack-2022-R48 & 77.2 & 7.54 & 6.743 & 1.588 \\
44 & Womack-2022-R57 & 79.8 & 6.57 & 6.855 & 1.715 \\
45 & Womack-2022-R63 & 82.3 & 6.53 & 6.404 & 1.75 \\
46 & Womack-2022-R70 & 82.0 & 7.34 & 5.955 & 1.74 \\
47 & Womack-2022-R78 & 77.6 & 5.4 & 5.255 & 1.716 \\
\hline
\end{tabular}
\label{table:full_data1}
\end{table}

\begin{table}
\vskip -0.2in
\centering
\caption{Surfaces, properties and wind-shade roughness model predictions (cont'd).}
\begin{tabular}{|c|c|c|c|c|c|c|}
\hline
Case \# &  Dataset name & $k_{\rm rms}^+$ & $k_{\rm s-data}/k_{\rm rms}$ & $k_{\rm s-mod}/k_{\rm rms}$ & ${\cal W}_{\rm L} \times 100$ \\ \hline
48 & Womack-2022-S10 & 34.9 & 4.32 & 3.013 & 0.55 \\
49 & Womack-2022-S17 & 47.9 & 5.97 & 5.472 & 0.97 \\
50 & Womack-2022-S39 & 67.0 & 8.62 & 6.344 & 1.414 \\
51 & Womack-2022-S48 & 71.0 & 7.89 & 6.443 & 1.553 \\
52 & Womack-2022-S57 & 73.8 & 6.77 & 6.25 & 1.604 \\
53 & Womack-2022-S63 & 72.6 & 6.57 & 6.058 & 1.625 \\
54 & Womack-2022-S70 & 71.3 & 5.75 & 5.693 & 1.606 \\
55 & Womack-2022-S78 & 71.7 & 6.71 & 5.317 & 1.581 \\
56 & Chung-2024-0p09 & 33.6 & 2.6 & 1.179 & 0.519 \\
57 & Chung-2024-0p18 & 47.0 & 5.18 & 3.537 & 1.13 \\
58 & Chung-2024-0p36 & 47.0 & 7.23 & 5.569 & 1.58 \\
59 & Forooghi-2017-A7088 & 22.5 & 7.2 & 10.363 & 2.141 \\
60 & Forooghi-2017-A7060 & 22.5 & 6.58 & 7.714 & 1.683 \\
61 & Forooghi-2017-A7040 & 22.5 & 5.49 & 4.81 & 1.175 \\
62 & Forooghi-2017-A7030 & 22.5 & 4.55 & 3.222 & 0.883 \\
63 & Forooghi-2017-A7020 & 22.5 & 2.57 & 1.447 & 0.5 \\
64 & Forooghi-2017-A3588 & 22.5 & 6.3 & 10.226 & 2.188 \\
65 & Forooghi-2017-A1588 & 22.5 & 9.04 & 10.277 & 2.271 \\
66 & Forooghi-2017-A0088 & 22.5 & 9.58 & 10.462 & 2.336 \\
67 & Forooghi-2017-A0060 & 22.5 & 9.2 & 7.517 & 1.786 \\
68 & Forooghi-2017-A0040 & 22.5 & 7.89 & 4.671 & 1.242 \\
69 & Forooghi-2017-A0030 & 22.5 & 6.47 & 2.869 & 0.878 \\
70 & Forooghi-2017-A0020 & 22.5 & 3.29 & 0.961 & 0.398 \\
71 & Forooghi-2017-B7088 & 22.5 & 4.2 & 7.441 & 1.841 \\
72 & Forooghi-2017-B7060 & 22.5 & 3.85 & 5.277 & 1.443 \\
73 & Forooghi-2017-B7040 & 22.5 & 3.1 & 3.871 & 1.124 \\
74 & Forooghi-2017-B7030 & 22.5 & 2.52 & 2.542 & 0.849 \\
75 & Forooghi-2017-B7020 & 22.5 & 1.67 & 1.365 & 0.532 \\
76 & Forooghi-2017-B3588 & 22.5 & 4.97 & 7.49 & 1.944 \\
77 & Forooghi-2017-B1588 & 22.5 & 5.44 & 7.189 & 1.951 \\
78 & Forooghi-2017-B0088 & 22.5 & 5.75 & 7.0 & 1.991 \\
79 & Forooghi-2017-C7088 & 22.5 & 9.14 & 12.416 & 2.373 \\
80 & Forooghi-2017-C7060 & 22.5 & 8.51 & 9.781 & 1.933 \\
81 & Forooghi-2017-C7040 & 22.5 & 7.05 & 3.871 & 1.124 \\
82 & Forooghi-2017-C7030 & 22.5 & 5.67 & 3.947 & 1.004 \\
83 & Forooghi-2017-C7020 & 22.5 & 3.47 & 1.847 & 0.591 \\
84 & Forooghi-2017-C3588 & 22.5 & 9.94 & 12.583 & 2.444 \\
85 & Forooghi-2017-C1588 & 22.5 & 10.88 & 12.541 & 2.487 \\
86 & Forooghi-2017-C0088 & 22.5 & 11.78 & 12.716 & 2.508 \\
87 & Forooghi-2017-D7088 & 22.5 & 4.96 & 8.95 & 2.082 \\
88 & Forooghi-2017-D0088 & 22.5 & 6.75 & 8.475 & 2.17 \\
89 & Forooghi-2017-D0088s & 22.5 & 7.47 & 8.453 & 2.164 \\
90 & Forooghi-2017-D0088a & 22.5 & 5.08 & 3.273 & 1.088 \\
91 & Barros-2018-pm15-Tile4 & 14.4 & 0.52 & 0.491 & 0.184 \\
92 & Barros-2018-pm10-Tile4 & 11.8 & 1.077 & 0.686 & 0.219 \\
93 & Barros-2018-pm05-Tile4 & 10.4 & 2.046 & 1.224 & 0.376 \\
94 & Yang-2021-lam25 & 36.6 & 2.105 & 5.81 & 1.39 \\
95 & Yang-2021-lam50 & 41.6 & 1.12 & 2.716 & 1.072 \\
96 & Yang-2021-lam60 & 40.3 & 0.736 & 1.705 & 0.844 \\
97 & Yang-2021-lam70 & 37.2 & 0.423 & 0.962 & 0.599 \\
98 & Yang-2021-lam80 & 31.8 & 0.182 & 0.492 & 0.35 \\
99 & Yang-2021-lam90 & 23.1 & 0.18 & 0.364 & 0.145 \\
100 & Medjnoun-2021-Iter1 & 189.7 & 2.27 & 3.99 & 0.874 \\
101 & Medjnoun-2021-Iter12 & 111.0 & 4.15 & 10.475 & 1.71 \\
102 & Medjnoun-2021-Iter13 & 182.0 & 2.78 & 8.3 & 1.466 \\
103 & Medjnoun-2021-Iter123 & 128.8 & 4.89 & 11.934 & 1.934 \\
\hline
\end{tabular}
\label{table:full_data2}
\end{table}

\vfill
\newpage

\section*{Appendix B: Notebook to compute sandgrain roughness}

The calculation method of wind-shade factor and sandgrain roughness for a given surface with a known height function $h(x,y)$, together with sample data, are provided via a notebook at 
\filelinknotebook .
It computes the wind-shade factor 
${\cal W}_{\rm L}$ and sandgrain roughness for two different rough surfaces. The elevation map is read from a {\it *.mat} file as a real array (typically for $n_x \times n_y \sim O(100^2)-O(1000^2)$ pixels). First, the backward horizon function and backward horizon angle $\beta$ are determined separately for each line along the  $x$-direction. If the flow is expected to be in a different direction, the elevation map should first be rotated so that $x$ becomes the flow direction. For some of the surfaces considered in the paper (e.g. wall attached cubes with significant sheltering), periodic boundary conditions are needed in order to capture effects of roughness elements upstream of the domain entrance. To capture this effect, the surface elevation map is replicated from the original length $n_x$ to a length $2n_x$ and the horizon function is determined for the entire double length. The horizon function and angle for the downstream half is then remapped to the original array of length $n_x$. For simplicity, we implement this approach as default for all surfaces considered. 

To evaluate $\partial h/\partial x$ and $\partial h/\partial y$ numerically, we use centered second order finite differencing, applied directly on the surface elevation map read from the database. For applications in which such data are expected to contain significant amounts of noise, filtering can be applied. 
Here, we treat the raw data without any filtering. Calculation of $\alpha$ and $\hat{n}_x$ then proceeds for each of the points and the product is multiplied by the slope and shading function. 
The sheltering function is computed based on the difference between the specified angle $\theta$ and the local angle $\beta(x,y)$ according to Eq. \ref{eq:defshadefunc}.  For in-between locations, i.e. for discrete positions $x_i$ such that $F^{\rm sh}(x_{i-1},y)=0$ and $F^{\rm sh}(x_{i+1},y)=1$ or vice-versa, $F^{\rm sh}(x_{i},y)$ is set to a fractional value between 0 and 1, proportional to the exposed fraction (i.e. proportional to the difference between $\tan(\theta)$ and $\tan(\beta)$).

The full product as defined in Eq. \ref{eq:defwindshade} is averaged  over the entire surface. Then the sandgrain roughness length $k_s$ and roughness function $\Delta U^+$ are computed. 

\begin{figure}
\centering
    \begin{subfigure}{0.6\textwidth}
        \includegraphics[width=1.15\linewidth]{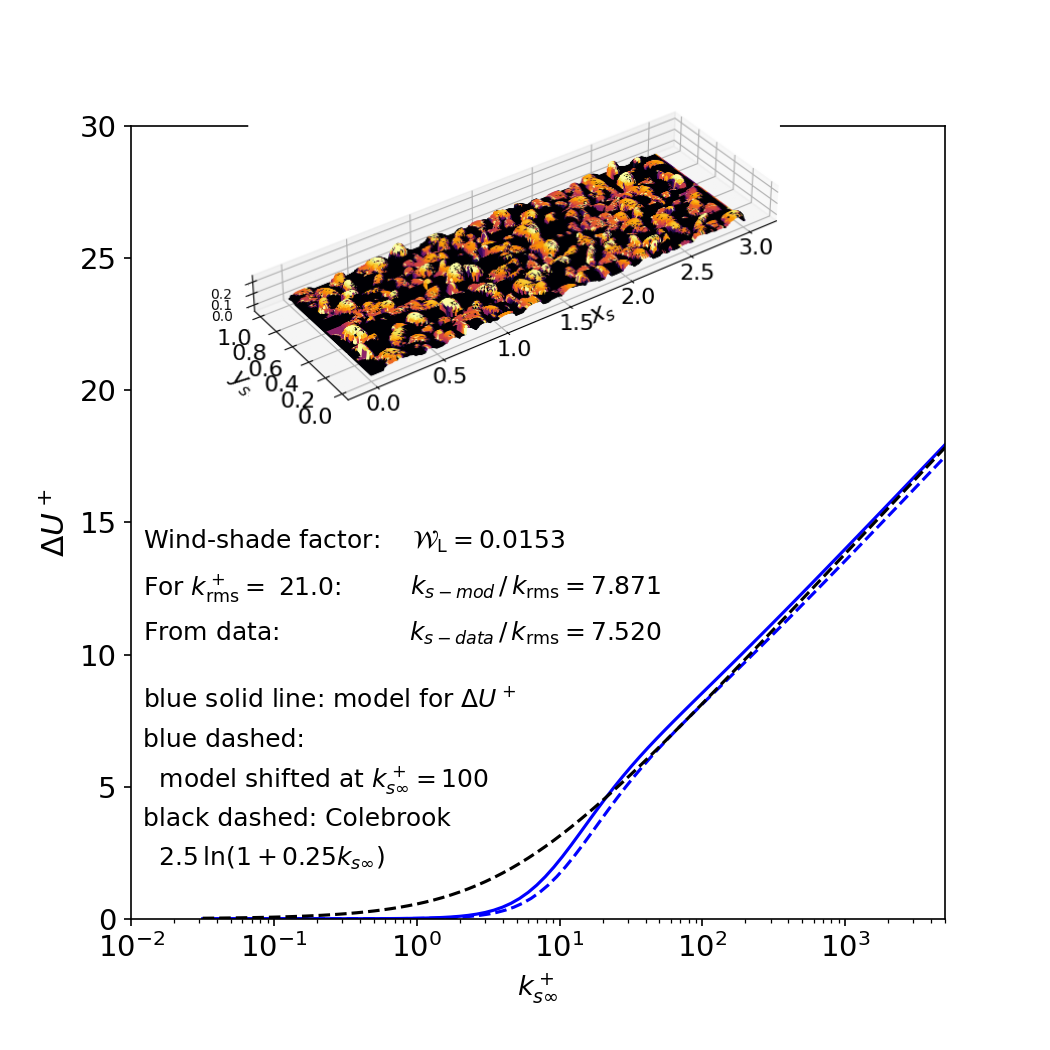}
        \vskip -0.2in
        \caption{ }
    \end{subfigure} 
     \begin{subfigure}{0.6\textwidth}
        \includegraphics[width=1.15\linewidth]{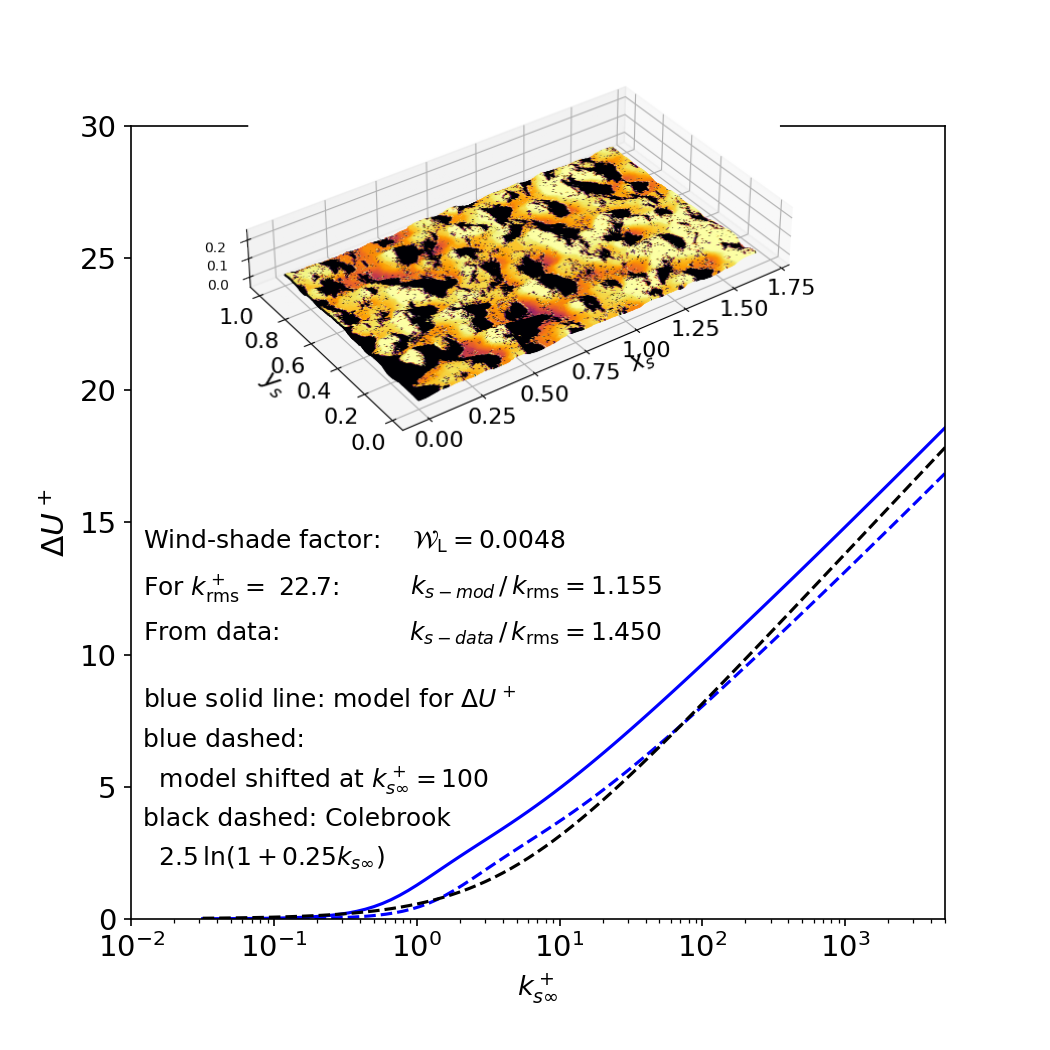}
        \vskip -0.2in
        \caption{ }
    \end{subfigure}  
   \caption{Two sample outputs from executing the notebook that can be found in \filelinknotebook. It computes the wind-shade factor ${\cal W}_{\rm L}$ and sandgrain roughness for two different rough surfaces. In (a) is shown the case C19 and in (b) the case Data1 (Tile 1) from the data of \cite{flack2020skin} available in the roughness database.}
   \vspace{-0.1in}
\label{fig:notebookfigs}
\end{figure}

The notebook at \filelinknotebook  is applied to two sample surfaces obtained from the roughness database 
(\href{https://roughnessdatabase.org}{\color{blue}https://roughnessdatabase.org}).
The outputs of the notebook applied to these surfaces are shown in Fig. \ref{fig:notebookfigs}. In (a) the surface C19 from \cite{jouybari2021data} is analyzed, while in  (b) the case ``Data1 (Tile 1)'' from the experiments of \cite{flack2020skin} is shown. For the notebook demonstration applied to the latter surface, only a smaller subset of the surface elevation map (500$\times$ 300 pixels) was used to compute ${\cal W}_{\rm L}=0.48$, whereas for the values listed in Table \ref{table:full_data1}, evaluation of ${\cal W}_{\rm L}$ was done over a larger map subset consisting of 1024$\times$ 512 pixels (as shown in Fig. \ref{fig:allfigs} e), and leading to a slightly larger value of 
${\cal W}_{\rm L}=0.50$. 

\vfill
\newpage

\section*{Appendix C: Analytical evaluations for aligned and staggered wall\dc{-}attached cubes}

\begin{figure}
  \centerline{\includegraphics[width=9cm]{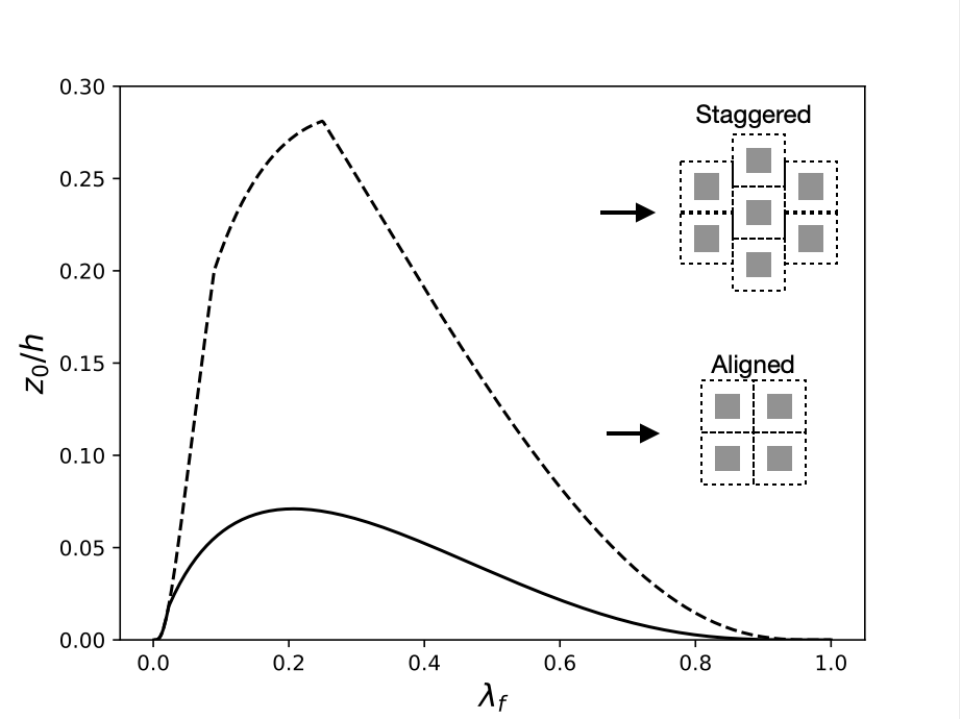}}
  \vspace{-0.1in}
  \caption{Roughness length normalized by cube height as function of frontal (or planform) area fraction $\lambda_f$, as predicted by the wind shade roughness model. The solid line is for aligned cubes and dashed line for staggered arrangement. The default parameters $a_p=3$, $\theta=10^o$, and $p=8$ are used.}
   \vspace{-0.1in}
\label{fig:z0vslambda}
\end{figure}

For the case of aligned cubes, the wind shade factor ${\cal W}_{\rm L}$ can be computed analytically as follows. For cubes, the planform and frontal solidities are equal, $\lambda_p=\lambda_f$. Assuming cube height $h$, the gap distance between the cubes in both $x$ and $y$ directions is $d=h \,(1/\sqrt{\lambda_p}-1)$. The exposed height is then $h \, \min\{1,(1/\sqrt{\lambda_p}-1) \tan \theta\}$. On this segment, the average pressure of ramp flow becomes stagnation point flow with $\alpha = \pi/2$ so that $\alpha/(\alpha+\pi) = 1/3$. Since $\hat{n}_x=1$, the average pressure is $U_k^2/3$. The slope $\partial h/\partial x$ is a delta function on these surfaces and the integral is simply the frontal exposed area times the averaged pressure. The average over reference planform area thus includes the probability $\lambda_p$ and the resulting wind shade factor is given by 

\be {\cal W}_{\rm L} = \frac{1}{3} \, \lambda_f \, \min\{1,(1/\sqrt{\lambda_f}-1) \tan \theta \},
\ee
(since $\lambda_f=\lambda_p$).  The reference scale $k_p^\prime$ is obtained by recognizing that   $ \langle h \rangle = h \lambda_p = h \lambda_f$  and the positive height fluctuations above the mean are $h'=h-h \lambda_f = h (1- \lambda_f)$, with probability $\lambda_f$. 
Then, 
\be k_p^\prime  =  h \, (1-\lambda_f) \, \lambda_f^{1/p},
\ee
and finally:
\be
\frac{z_0}{h} =   {a_p} \,  (1-\lambda_f)\, \lambda_f^{1/p}   \, \,\exp\left(-\kappa \,\, \left[\frac{1}{3}  \lambda_f\, \min\{1, \, (1/\sqrt{\lambda_f}-1) \tan \theta \} \right] ^{-1/2} \right).
\ee
 Figure \ref{fig:z0vslambda} (solid line) shows the resulting roughness length as function of $\lambda_f$ for $\theta=0.175$ (10 degrees), $a_p=3$, and $p=8$. Its peak is near $\lambda_f \approx 0.2$ with $z_0/h \approx 0.07$, which agrees qualitatively with available datasets \cite{Leonardi2010,Hagishima2009,Cheng2007,Hall1996,yang2016exponential} and the shading model of \cite{yang2016exponential}, and is quantitatively within the considerable empirical spread of the available data. 

Similar derivation for equispaced staggered cube arrays leads to the conclusion that for $\lambda_f=\lambda_p \leq 1/4$ (large spacing), the distance between shaded cubes is now $h[1+2(1/\sqrt{\lambda_f}-1)]=
h\,(2/\sqrt{\lambda_f}-1)$ and the exposed height is $h \, \min\{1,(2/\sqrt{\lambda_f}-1) \tan \theta\}$. The exposed area fraction (with respect to planform total area) is then equal to $\lambda_f \, \min\{1,(2/\sqrt{\lambda_f}-1) \tan \theta\} $.
For $\lambda_f > 1/4$ (close spacing), two distances exist: 
neighboring cubes (shifted) at immediate distance $d=h(1/\sqrt{\lambda_f}-1)$ and exposed width equal to $(h-d)$. And cubes at larger distance $h\,(2/\sqrt{\lambda_f}-1)$ with exposed width $d$. 
The resulting total exposed area fraction is therefore 
$[\, h\,(h-d)\,(1/\sqrt{\lambda_f}-1) \tan \theta 
+h \,d\,(2/\sqrt{\lambda_f}-1) \tan\theta \, ]/(h+d)^2 = 
 \lambda_f(\lambda_f^{-1}-1) \tan\theta $. 
The resulting wind shade factor can then be written as 
\be
{\cal W}_{\rm L} = \frac{1}{3} \, \lambda_f \, \min\{1,(2/\sqrt{\lambda_f}-1)\tan\theta,(\lambda_f^{-1}-1)\tan\theta\}.
\ee
The roughness length is then given again by
\be
\frac{z_0}{h} =   {a_p} \, (1-\lambda_f)\, \lambda_f^{1/p}   \, \,\exp\left(-\kappa \,\, {\cal W}_{\rm L}^{-1/2} \right).
\ee

The dashed line in Fig. \ref{fig:z0vslambda} shows the prediction for staggered cubes, for the same $a_p$, $\theta$ and $p$ parameters. It is noteworthy that in this case $z_0$ appears somewhat overpredicted compared to the available data as well as compared to the sheltering model of \cite{yang2016exponential}. The latter also includes sideways expansion of the sheltered region, thus decreasing drag for staggered cubes. The current version of the model only includes streamwise sheltering thus exposing more of the downstream cubes to flow compared to the prior model of \cite{yang2016exponential}.

\bibliographystyle{jfm}
\bibliography{roughness-biblio}

\end{document}